\documentclass[fleqn,usenatbib,useAMS]{mnras}

\usepackage{newtxtext,newtxmath}

\usepackage[T1]{fontenc}

\DeclareRobustCommand{\VAN}[3]{#2}
\let\VANthebibliography\thebibliography
\def\thebibliography{\DeclareRobustCommand{\VAN}[3]{##3}\VANthebibliography}

\usepackage{graphicx}	
\usepackage{amsmath}	
\usepackage{amssymb}	
\usepackage{capt-of}    
\defcitealias{Wang16}{W16}
\defcitealias{Swartz11}{S11}

\usepackage{xcolor}

\newcommand{\ergs}{\ensuremath{\,\text{erg}\,\text{s}^{-1}}}
\newcommand{\fluxunits}{\ensuremath{\rm \,erg\,s^{-1}\,cm^{-2}}}
\newcommand{\ulxlimit}{\ensuremath{10^{39}\ergs}}
\newcommand{\sfrunit}{\ensuremath{\rm\,M_\odot\,yr^{-1}}}
\newcommand{\massunit}{\ensuremath{\rm\,M_\odot}}
\newcommand{\Mpc}{\ensuremath{\,\text{Mpc}}}
\newcommand{\mass}{\ensuremath{M_{\star}}}
\newcommand{\aunits}{\ensuremath{\rm\,\text{ULXs per } M_\odot\,yr^{-1}}}
\newcommand{\bunits}{\ensuremath{\rm\,\text{ULXs per } 10^{12} M_\odot}}
\newcommand{\asec}{\ensuremath{\,\text{arcsec}}}

\newcommand{\good}{\lq{}reliable\rq{}}
\newcommand{\bad}{\lq{}unreliable\rq{}}

\newcommand{\full}{\lq{}full\rq{}}
\newcommand{\non}{\lq{}non-AGN\rq{}}

\newcommand{\CSC}{\textit{CSC 2.0}}
\newcommand{\hec}{\textit{HECATE}}
\newcommand{\chandra}{\textit{Chandra}}

\newcommand{\hyper}{\textit{HyperLEDA}}
\newcommand{\ned}{\textit{NED}}
\newcommand{\nedd}{\textit{NED-D}}
\newcommand{\twomass}{\textit{2MASS}}
\newcommand{\sdss}{\textit{SDSS}}
\newcommand{\iras}{\textit{IRAS}}
\newcommand{\wise}{\textit{WISE}}

\title[A census of ULXs in the local Universe]{A census of ultraluminous X-ray sources in the local Universe}

\author[K. Kovlakas, A. Zezas, J. Andrews et al.]{%
K. Kovlakas,$^{1,2}$\thanks{e-mail: kkovlakas@physics.uoc.gr}
A. Zezas,$^{1,2,3}$
J. J. Andrews,$^{4,5}$
A. Basu-Zych,$^{6,7}$
T. Fragos,$^8$
\newauthor
A. Hornschemeier,$^{6,9}$
B. Lehmer,$^{10}$ and
A. Ptak$^{6}$
\\
$^1$Physics Department, University of Crete, GR 71003, Heraklion, Greece\\
$^2$Institute of Astrophysics, Foundation for Research and Technology-Hellas, GR 71110 Heraklion, Greece\\
$^3$Harvard-Smithsonian Center for Astrophysics, 60 Garden Street, Cambridge, MA 02138, USA\\
$^4$CIERA, Northwestern University, 1800 Sherman Ave, Evanston, IL 60201, USA\\
$^5$Department of Physics and Astronomy, Northwestern University, 2145 Sheridan Rd, Evanston, IL 60208, USA \\
$^6$NASA Goddard Space Flight Center, Laboratory for X-ray Astrophysics, Greenbelt, MD 20771, USA\\
$^7$Department of Physics, University of Maryland Baltimore County, Baltimore, MD 21250, USA\\
$^8$Geneva Observatory, University of Geneva, Chemin des Maillettes 51, 1290 Sauverny, Switzerland\\
$^9$The Johns Hopkins University, Homewood Campus, Baltimore, MD 21218, USA\\
$^{10}$Department of Physics, University of Arkansas, 825 West Dickson Street, Fayetteville, AR 72701, USA
}

\date{Accepted XXX. Received YYY; in original form ZZZ}

\pubyear{2020}

\begin{document}
\label{firstpage}
\pagerange{\pageref{firstpage}--\pageref{lastpage}}
\maketitle

\begin{abstract}
Using the {\it Chandra Source Catalog 2.0} and a newly compiled catalogue of galaxies in the local Universe, we deliver a census of ultraluminous X-ray source (ULX) populations in nearby galaxies. We find 629 ULX candidates in 309 galaxies with distance smaller than 40\,Mpc. The foreground/background contamination is ${\sim}20\%$. The ULX populations in bona-fide star-forming galaxies scale on average with star-formation rate (SFR) and stellar mass ($M_\star$) such that the number of ULXs per galaxy is $0.45^{+0.06}_{-0.09}\times\frac{\rm SFR}{\rm M_\odot\,yr^{-1}}{+}3.3^{+3.8}_{-3.2}\times\frac{M_\star}{\rm M_\odot}$. The scaling depends strongly on the morphological type. This analysis shows that early spiral galaxies contain an additional population of ULXs that scales with $M_\star$. We also confirm the strong anti-correlation of the ULX rate with the host galaxy's metallicity. In the case of early-type galaxies we find that there is a non-linear dependence of the number of ULXs with $M_\star$, which is interpreted as the result of star-formation history  differences. Taking into account age and metallicity effects, we find that the predictions from X-ray binary population synthesis models are consistent with the observed ULX rates in early-type galaxies, as well as, spiral/irregular galaxies.
\end{abstract}



\section{Introduction}

Ultraluminous X-ray sources (ULXs) are galactic point-like X-ray sources, not associated with an active galactic nucleus, with X-ray luminosities above the Eddington limit of an accreting stellar-mass black hole (${\gtrsim}10^{39}\ergs$; for a recent review see \citealt{Kaaret17}). Soon after their discovery by the {\it Einstein} observatory \citep{Long83,Fabbiano89}, three scenarios were proposed to explain their high luminosities. Initially, it was proposed that ULXs are accreting black holes (BHs) with masses in the range between stellar-mass and supermassive BHs (${\sim}10^2{-}10^6\rm\,M_\odot$), i.e., intermediate-mass BHs (IMBHs; \citealt{Colbert99}; \citealt{Makishima00}; \citealt{Marel04}). This scenario was dismissed on theoretical grounds due to difficulties in the formation of X-ray binaries with IMBHs \citep[e.g.,][]{Kuranov07}, although a few cases are still viable (e.g., ESO 243-49 HLX-1: \citealt{Farrell09}; M82 X-1: \citealt{Ptak99}). The second scenario involves stellar-mass BHs (with masses in the range of Galactic BHs, ${\lesssim}15\rm\,M_\odot$; \citealt{Remillard06}), which may have super-Eddington luminosities when accreting at super-critical rates \citep[e.g.,][]{Begelman02}.
In the third scenario the ULX luminosities are the result of geometrical beaming of the emitted radiation \citep{King01} due to the formation of a funnel in the central part of the supercritical accretion disk \citep[e.g.,][]{Abramowicz88,Sadowski14}.

The combination of these two scenarios can explain the observed ULX population with $L_{\rm X}{<}10^{41}\ergs$ as the high-luminosity end of the luminosity function of X-ray binaries (XRBs). Recently, the discovery of pulsating ULXs \citep{Bachetti14,Furst16,IsraelA,IsraelB,Carpano18} showed that the accretor can even be a neutron star (NS), making the super-Eddington accretion scenario necessary for their explanation (e.g., \citealt{Fragos15,King16,King17,Middleton17,Misra20}).

The above three scenarios highlight the importance of ULXs in understanding massive binary evolution and accretion physics at extreme accretion rates. The latter is crucial for shedding light at the formation of compact object mergers that are detected as short gamma-ray bursts and gravitational wave sources \citep{Berger14,Finke17,Marchant17,Mondal20}. In addition, the extreme emission of ULXs may have played a role in the heating of the Universe during the epoch of reionization (e.g., \citealt{Venkatesan01,Madau04}; however see \citealt{Das17,Madau17}).

\label{txt:introhost}

A deeper understanding of ULXs can be obtained by detailed spectral and timing studies of individual sources \citep[e.g.,][]{Gladstone09,Middleton15,Walton19,Koliopanos19}. While these studies provide valuable insights into the physics and nature of the accretion, they offer limited information on the formation and evolution pathways of ULXs. The latter can be better constrained by identifying their optical counterparts and/or studying their populations in the context of their host galaxies. Since ULXs are rare and usually found in distant galaxies, the identification of optical counterparts and measurement of the compact object masses are observationally challenging \citep{Angelini01,Colbert02,Swartz04,Feng08}. Consequently, ULX demographics and scaling relations between the ULX content and stellar population parameters of their host galaxies, such as SFR and \mass{}, are important tools for understanding the nature and evolution of ULXs via the comparison with binary population synthesis models \citep[e.g.,][]{Rappaport05,Wiktorowicz17}.

Early surveys of nearby galaxies revealed an overabundance of ULXs in late-type galaxies (LTGs) \citep[e.g.,][]{Roberts00}, while direct association of ULXs with star-forming regions of their hosts {connected ULXs} with young stellar populations, indicating that the majority of ULXs are a subset of high-mass X-ray binaries \citep[HMXBs; e.g.,][]{Fabbiano01,Roberts02,Gao03,Zezas07,Wolter04,Kaaret04,Anastasopoulou16,Wolter18}. Nevertheless, a small but significant fraction of ULXs are found in early-type galaxies, and therefore are connected to old stellar populations, i.e. ultraluminous low-mass X-ray binaries (LMXBs; \citealt{Angelini01,Colbert02,Swartz04,Kim04,Fabbiano06,Feng08}.
These demographic studies agree on two findings:
\begin{itemize}
    \item[a)] Dwarf galaxies have been found to host more ULXs than expected given their SFR \citep{Swartz08,Walton11,Plotkin14,Tzanavaris16}.
    \item[b)] An observed excess of ULXs (and XRBs in general) in low-metallicity galaxies \citep[e.g.,][]{Mapelli10,Prestwich13,Brorby14,Douna15,Basu16}
\end{itemize}

The excess in low-metallicty galaxies has highlighted the effect of metallicity on the accretor's mass and the evolutionary paths of ULXs \citep{Heger03,Soria05,Belczynski10,Linden10,Mapelli11,Marchant17}. The same effect has been invoked to interpret the X-ray emission properties of high-redshift galaxies (Lyman Break Galaxies and Lyman Break Analogs; \citealt{Basu13a,Basu13b,Basu16,Brorby16,Lehmer16}), as demonstrated by binary population synthesis  models \citep{Linden10,Fragos13a,Fragos13b,Wiktorowicz17}.

In the era of {\it ROSAT} and the early days of \chandra{}, ULX demographics were limited to a few tens of sources and galaxies \citep[e.g.,][]{Colbert02,Swartz04,Liu06}. Therefore, these studies were unable to resolve the dependence of ULX populations on the stellar populations of their host galaxies. The first quantitative study of the rate of ULXs in the local Universe, based on a complete sample of galaxies up to $14.5\Mpc{}$, showed that the observed population of ULXs is \lq{}consistent with the extrapolation of the luminosity function of ordinary X-ray binaries\rq{} (LMXBs and HMXBs in early- and late-type galaxies respectively; \citealt{Swartz11}). However, the volume limit resulted into an oversampling of irregular galaxies and under-representation of elliptical galaxies. The largest to date demographic study of ULXs (343 galaxies) was presented in \citet{Wang16}, using \chandra{} observations until 2007. This work constrained the X-ray luminosity function (XLF) parameters of ULXs in galaxies of different morphological types, and showed that elliptical galaxies host more ULXs than in samples of previous studies. However, \citet{Wang16} focused on XLFs of ULXs and did not study their scaling with the SFR, \mass{} and metallicity of their hosts.
The most recent catalogue of ULX candidates was presented in \citet{Earnshaw19}. It includes 384 ULXs drawn from the 3XMM-DR4 catalogue. This study showed that the hardness ratio (HR) distribution of ULXs is similar to that of the lower-luminosity XRBs, but not AGN, and mostly independent of the environment (elliptical vs. spiral galaxies). However, this study focused on the X-ray spectral and timing properties of the sources rather than their connection to their hosts.

The {\it Chandra Source Catalog 2.0} (\CSC{}) gives a unique opportunity to study the demographics of ULXs in the context of the stellar populations of their host galaxies (SFR, \mass{}, metallicity) by utilising the largest available sample of X-ray sources, and a new catalogue of galaxies in the local Universe.

This paper is organised as follows: in Sections \ref{txt:galaxysample} and \ref{txt:xraysample} we describe the sample of host galaxies and X-ray sources, respectively. In Section \ref{txt:results} we report the results on ULX demographics and their connection with stellar population parameters, while in Section \ref{txt:discussion} we discuss the implications of this study in comparison to previous studies and ULX population models. Finally, in Section \ref{txt:summary} we summarise the main findings. Unless stated otherwise, the reported uncertainties correspond to $68\%$ confidence intervals.

\begin{figure*}
    \centering
    \includegraphics[width=0.95\textwidth]{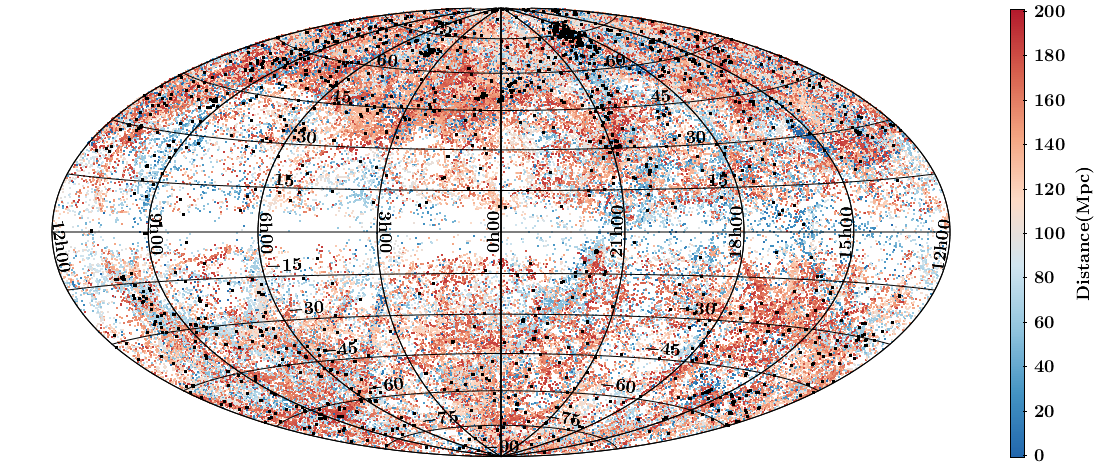}
    \caption{Sky map of the \hec{} galaxies in Galactic coordinates with colour denoting their distance. Galaxies included in the \CSC{} are shown as black points. Note the sparsity of sources in the plane of the Milky Way (Zone of Avoidance; ZOA) and the increased density in the North and West parts due to the inclusion of \sdss{} galaxies in the \hyper{}.}
    \label{fig:allsky}
\end{figure*}

\section{The galaxy sample}
\label{txt:galaxysample}

\begin{table*}
\caption{Parameters of the 2218 host galaxies. Only a small portion of this table is shown here, indicative of the various cases (e.g., flags, missing parameters). The full table is available in the online journal.}
\label{tbl:hosts}
\begin{tabular}
    {l@{\hskip 6pt}
     l@{\hskip 6pt}
     r@{\hskip 6pt}
     r@{\hskip 6pt}
     r@{\hskip 6pt}
     r@{\hskip 6pt}
     r@{\hskip 6pt}
     c@{\hskip 6pt}
     r@{\hskip 6pt}
     r@{\hskip 6pt}
     r@{\hskip 6pt}
     c@{\hskip 6pt}
     c@{\hskip 6pt}
     c@{\hskip 6pt}
     c@{\hskip 6pt}
     c@{\hskip 6pt}
     c@{\hskip 6pt}
     c}
\hline
PGC & ID & $\alpha$ & $\delta$ & $R_1$ & $R_2$ & $\phi$ & $T$ & $D$ & $\log{\rm SFR}$ & $\log\mass{}$ & $Z$ & AGN & $f_{25}$ & U & $N_{\rm obs}$ & $N_{\rm f/b}$ & $N_{\rm ulx}$ \\
(1) & (2) & (3) & (4) & (5) & (6) & (7) & (8) & (9) & (10) & (11) & (12) & (13) & (14) & (15) & (16) & (17) & (18) \\\hline
2557 & NGC0224 & 10.684684 & 41.268978 & 88.91 & 34.83 & 35 & 3.0$\pm$0.4 & 0.8 & -0.33 & 10.61 &  & Y & 0.46 &  & 0 & 0.01 & $0.00_{-0.00}^{+1.13}$\\[1pt]
16570 & NGC1741B & 75.398106 & -4.263220 & 0.46 & 0.23 & 42 & 6.8$\pm$3.3 & 55.7 &  & 9.34 &  &  & 1.00 & * & 2 & 0.09 & $1.91_{-1.13}^{+1.85}$\\[1pt]
23324 & UGC04305 & 124.768125 & 70.721674 & 3.96 & 2.79 & 15 & 9.9$\pm$0.5 & 3.4 & -2.07 & 8.77 &  &  & 0.27 & * & 1 & 0.00 & $1.00_{-0.73}^{+1.49}$\\[1pt]
35249 & NGC3683 & 171.882672 & 56.877021 & 0.87 & 0.35 & 124 & 4.8$\pm$0.7 & 33.3 & 0.94 & 10.81 & 8.76 & Y & 1.00 & * & 5 & 0.12 & $4.88_{-1.93}^{+2.62}$\\[1pt]
38742 & NGC4150 & 182.640252 & 30.401578 & 0.99 & 0.66 & 148 & -2.1$\pm$0.7 & 13.6 & -0.84 & 9.90 &  & N & 1.00 & * & 2 & 0.02 & $1.98_{-1.14}^{+1.84}$\\
\hline
\end{tabular}

\begin{minipage}{0.98\linewidth}    
Columns description:
(1) identification number in the \hyper{} and the \hec{};
(2) galaxy name;
(3), (4) right ascension and declination (J2000.0) ($\degr$);
(5)-(7) the semi-major and -minor axes (\arcmin{}), and the North-to-East position angle (\degr{});
(8) morphological code, $T$ (see \autoref{tab:rc2});
(9) distance (Mpc);
(10) decimal logarithm of SFR [$\sfrunit{}$];
(11) decimal logarithm of \mass{} [$\massunit$];
(12) metallicity ($12+\log\left(O/H\right)$);
(13) the galaxy hosts an AGN;
(14) fraction of $D_{25}$ covered by the \CSC{} stacks;
(15) * if the galaxy is used in the analysis in this paper (see \S\ref{txt:fieldofview});
(16) number of observed sources with $L_X{>}10^{39}\ergs{}$, excluding nuclear sources if the host is classified as AGN (see \S\ref{txt:interlopersAGN});
(17) number of expected foreground/background source contamination in the ULX regime;
(18) number of ULXs by subtracting interlopers.
Columns (1)-(13) are taken from the \hec{}, while the rest are described in \S\ref{txt:xraysample}.
\end{minipage}
\end{table*}

We use the \textit{Heraklion Extragalactic Catalogue} (\hec{}), a compilation of all galaxies
within $200\Mpc$, from the \hyper{} \citep{Makarov14}, arguably the most complete compilation of galaxies with homogenised parameters. The \hec{} adopts positions, sizes, morphological classifications, and redshifts from the \hyper{}. These are complemented with size and redshift information from other catalogues when not available in the \hyper{}. It also provides robust estimates of distances, along with SFRs, stellar masses, metallicities and nuclear activity classifications.
In the following paragraphs we provide a brief summary of the relevant properties of the catalogue. A detailed description of the catalogue and the data it contains is presented in Kovlakas et al. (in prep).

The \hec{} is based on all \hyper{} galaxies (object type \lq{}G\rq{}) with Virgo-infall corrected radial velocities less than $14000\rm\,km\,s^{-1}$ (corresponding to distances ${\lesssim}200\Mpc$ and redshifts ${\lesssim}0.047$). When redshift and size information (semi-major/minor axes and position angles) are not directly  available in the \hyper{}, they are obtained from other databases or catalogues (e.g., \ned{}, \sdss{}, \twomass)\footnote{None of these galaxies (with supplemented redshift/size information) is included in our analysis because they lack other required information (e.g., morphological classifications, IR photometry which is used for deriving SFR and stellar mass measurements).}.
\autoref{fig:allsky} shows the position of the galaxies in the \hec{} in Galactic coordinates.

The \hec{} provides redshift-independent distances (e.g., based on the Cepheids, RR Lyrae, Tully-Fisher, surface-brightness fluctuations, tip of the red-giant branch methods) for ${\sim}10\%$ of the galaxies obtained from the \nedd{} \citep{Steer17}.
When only one distance measurement is available, it is adopted as is. In the case of multiple distance measurements, a statistical estimate is made using a weighted Gaussian Mixture model, with weights that penalise uncertain or old measurements. Subsequently, these distances along with the radial velocities for the same galaxies are used to train a Kernel Regression model which is the used to predict the radial-velocity based distance (and its uncertainty based on the intrinsic scatter) for all the other objects. More details on the method for calculating the galaxy distances are given in Kovlakas et al. (in prep).

The \hec{} provides SFR estimates for galaxies with reliable mid- and far-infrared photometric measurements from the \iras{} and the \wise{}. Depending on the availability and quality of photometry, three different SFR indicators were computed based on \iras{} photometry: (i) total-infrared (TIR; 24, 60 and $100\umu$ calibrations of \citealt{Dale02} and \citealt{Kennicutt12}), (ii) far-infrared (FIR; 60 and $100\umu$ calibrations of \citealt{Helou88} and \citealt{Kennicutt98}), and (iii) $60\umu$ (calibrations of \citealt{Rowan99}). Additionally, \wise{} photometry, obtained from the forced photometry catalogue of \citet{Lang16} for galaxies in the \sdss{} footprint, is used to provide $12\umu$ and $22\umu$-based SFR estimates (calibrations of \citealt{Cluver17}). An \lq{}adopted\rq{} SFR for each galaxy is obtained by homogenising the SFR indicators (using the TIR-based one as reference) and selecting for each galaxy the first available SFR estimate in the following order of preference: TIR, FIR, $60\umu$, $12\umu$, $22\umu$.
It should be noted that H$\alpha$ SFR indicators are more appropriate than infrared (IR) indicators for studying the connection of ULXs with young stellar populations, since the latter probe star formation at scales (${\sim}100\,\rm Myr$), longer than the life-time of HMXBs (cf. \citealt{Kouroumpatzakis20}).
However, IR photometry is readily available for a significant fraction of our sample, and it is generally well correlated with H$\alpha$.

The integrated \twomass{} $K$-band photometry and \sdss{} $g{-}r$ colour, were used to estimate the stellar masses of the galaxies, using the mass-to-light ratio calibrations of \citet{Bell03}. For galaxies without \sdss{} photometry, the \hec{} assumes the mean mass-to-light ratio of the galaxies with \sdss{} data.

In addition, the \hec{} includes gas-phase metallicities based on \sdss{} spectroscopic data from the {\it MPA-JHU catalogue} \citep{Kauffmann03,Brinchmann04,Tremonti04}, using the \ion{O}{III}-\ion{N}{II} calibration in \citealt{PP04}. Based on the star-light subtracted \sdss{} spectra, the \hec{} identifies AGN on the basis of their location in optical emission-line ratio diagnostic diagram, using the multi-dimensional classification scheme of \citet{Stampoulis19}.

The Third Reference Catalogue of Bright Galaxies ({\it RC3}; \citealt{RC91}) has been the reference galaxy sample for several studies of ULXs \citep[e.g.][]{Swartz11,Wang16,Earnshaw19}. While it provides a wide range of information (positions, diameters, morphological types, photometry, and radial velocities), its small size (23022 galaxies) has been superseded by larger and more complete samples of galaxies.
The \hyper{}, and subsequently the \hec{}, provide a ${\sim}10$ times improvement in the sample size within our volume of interest ($D{<}200\,\Mpc{}$).
Therefore, the \hec{}, provides a much more complete census of the galaxy populations in the local Universe, supplemented by a wealth of additional information described in the previous paragraphs. This makes it more appropriate for the exploration of the multi-wavelength properties of galaxies based on serendipitous surveys.

\section{The X-ray sample}
\label{txt:xraysample}

To identify ULX candidates, we use the \CSC{}\footnote{\url{https://cxc.harvard.edu/csc2/}}, which is a publicly available catalogue of all the sources detected in \chandra{} observations performed up to the end of 2014. It contains 317167 X-ray sources, an improvement of more than a factor of $3$ compared to the previous version \citep[version 1.1;][]{Evans10}.

\subsection{Selection of sources}
\label{txt:xrayselection}

An X-ray source is associated with a \hec{} galaxy if it is located within its $D_{25}$ region.
The positional uncertainties of the sources are not considered, since they are negligible with respect to the dimensions of the galaxies: 95\% (98\%) of the sources have uncertainties less than the 1\% (10\%) of the semi-major axes of their host galaxies.
The few, galaxies without size information in the \hec{} (${\sim}2\%$ of the full sample) are excluded from the cross-matching.

Out of the 317167 sources in the \CSC{}, we associate 23043 sources to 2218 galaxies within a distance of 200\Mpc{}. The host galaxies are shown by black points in \autoref{fig:allsky}. The parameters of the host galaxies are listed in \autoref{tbl:hosts} (columns (1)-(13)), while the properties of the selected X-ray sources are given in \autoref{tbl:sources}.

\begin{table*}
\caption{Properties of the 23043 X-ray sources. Only a small portion of this table is shown here, indicative of the various cases (e.g., flags, missing parameters). The full table is available in the online journal.}
\label{tbl:sources}
\begin{tabular}{
    l@{\hskip 12pt}
    l@{\hskip 12pt}
    r@{\hskip 12pt}
    r@{\hskip 12pt}
    c@{\hskip 8pt}
    c@{\hskip 8pt}
    c@{\hskip 12pt}
    c@{\hskip 8pt}
    c@{\hskip 8pt}
    c@{\hskip 8pt}
    c@{\hskip 8pt}
    c@{\hskip 12pt}
    c@{\hskip 8pt}
    c@{\hskip 8pt}
    c}
\hline
PGC & ID & $\alpha$ & $\delta$ & $\log f$ & $\log f_{\rm lo}$ & $\log f_{\rm hi}$ & p & u & n & $c$ & $\log L_{\rm X}$ & $\log L_{\rm X, lo}$ & $\log L_{\rm X, hi}$ \\
(1) & (2) & (3) & (4) & (5) & (6) & (7) & (8) & (9) & (10) & (11) & (12) & (13) & (14) \\\hline
101 & 2CXO J000120.2+130641 & 0.33422 & 13.11141 &-14.37 & -14.54 & -14.25 &  &  &  & 0.14 & 39.44 & 39.28 & 39.56\\
1305 & 2CXO J002012.6+591501 & 5.05281 & 59.25038 &-13.36 & -13.40 & -13.32 &  & * &  & 0.97 & 36.47 & 36.42 & 36.50\\
2789 & 2CXO J004732.9-251748 & 11.88735 & -25.29692 &-12.15 & -12.16 & -12.15 & * &  &  & 0.17 & 39.02 & 39.01 & 39.03\\
12997 & 2CXO J032953.1-523054 & 52.47155 & -52.51524 &-13.87 & -13.97 & -13.80 &  &  & * & 0.02 & 40.60 & 40.51 & 40.68\\
42038 & 2CXO J123622.9+255844 & 189.09568 & 25.97891 & &  & -15.05 &  & * &  & 0.09 &  &  & 37.24 \\\hline
\end{tabular}

\begin{minipage}{0.98\linewidth}    
Columns description:
(1) identification number of host galaxy in the \hyper{} and the \hec{};
(2) name of master source in the \CSC{};
(3), (4) right ascension and declination (J2000.0) ($\degr$);
(5)-(7) decimal logarithm of flux (\lq{}flux\_aper90\_b\rq{}) and its 68\% confidence interval [$\fluxunits$];
(8) * if pileup source (lower limit on flux and luminosity);
(9) * if \bad{} source (see \S\ref{txt:xrayselection});
(10) * if nuclear source;
(11) galactocentric scale parameter;
(12-14) decimal logarithm of X-ray luminosity and its 68\% confidence interval [$\ergs{}$].
The data in columns (2)-(8) are taken directly from the \CSC{}, while those in columns (9)-(14) are described in \S\ref{txt:xraysample}.
\end{minipage}
\end{table*}

\label{txt:srcquality}
We characterise sources as \good{} or \bad{} (column (9) in \autoref{tbl:sources}) based on their attributes in the \CSC{}. A source is marked as \bad{} if any of the following conditions are met:
\begin{enumerate}
    \item the flux is zero (i.e., upper limit) or no confidence interval is provided,
    \item the \lq{}dither\_warning\_flag\rq{} is on: indicating that the highest peak of the power spectrum of the source occurs at the dither frequency (or its beat frequency) in all observations,
    \item the \lq{}streak\_src\_flag\rq{} is on: the source is found on an ACIS readout streak in all observations,
    \item the \lq{}sat\_src\_flag\rq{} is on: saturated in all observations.
\end{enumerate}
We find 3783 (16.4\%) \bad{} sources, out of which, 1040 (4.5\%) are characterised as such because of a flag, 1952 (8.5\%) have zero flux, and 791 (3.4\%) have missing confidence intervals.
The remaining 19260 sources (83.6\%) are characterised as \good{}. In the following analysis we consider only the \good{} sources. We note that the majority of the more luminous sources in our sample ($L_{\rm X}{\gtrsim}10^{41.5}\ergs{}$) are flagged as \bad{} (see \S\ref{txt:lumdist}), since they are more likely to be saturated.

\subsection{\chandra{} field-of-view coverage}
\label{txt:fieldofview}

The \chandra{} observations from which the X-ray sources in the \CSC{} are observed, typically target individual galaxies. The field of view is usually centred on the galaxy and covers fully its $D_{25}$ region. However, there are cases of large nearby galaxies that are partially covered, as well as, observations that target off-centre regions.

In order to measure the coverage of each galaxy by \chandra{}, we compute the fraction, $f_{25}$, of the $D_{25}$ region in the union of the stack-field-of-view\footnote{\url{http://cxc.harvard.edu/csc2/data_products/stack/fov3.html}} of all the stacks contributing in the \CSC{} (column (11) in \autoref{tbl:hosts}). We consider galaxies with $f_{25}>0.7$ as sufficiently covered. After visually inspecting multi-wavelength images of the galaxies without full coverage, we find that the missing area generally leaves a negligible fraction of the total SFR and \mass{} unaccounted for.

We find 34 galaxies (${<}2\%$) with coverage less than 70\%. Some galaxies may have poor coverage because of observations performed in sub-array mode (e.g., those focusing on known ULXs.)
Excluding these galaxies would bias our demographics; on the other hand, ULXs that are located in the unobserved area of the galaxies would also provide an incomplete picture of ULX populations. For this reason, we manually inspect for the presence of bright sources in {\it XMM-Newton} observations with wider field-of-view. Such observations are available for 16 objects, for which we find no other bright ($L_{\rm X}{>}10^{39}\ergs{}$) sources in their $D_{25}$ regions. Therefore, we include them in the following analysis since their ULX population is complete in our \chandra{}-based sample. The remaining 18 galaxies are excluded from the subsequent analysis (most of which are known to not host ULXs, e.g., SMC), but not from the provided catalogues.

\subsection{Survey coverage and representativeness}

\subsubsection{Source confusion}
\label{txt:disteffect}

At large distances, source confusion severely limits X-ray binary population studies. This effect is more prominent in studies of young stellar populations (such as ULXs; e.g., \citealt{Anastasopoulou16,Basu16}) due to the clumpy nature of star-forming regions \citep[e.g.,][]{Elmegreen96,Sun18}. Specifically, at $D{>}40\Mpc{}$, the half-arcsecond beam of \chandra{} is comparable to the angular sizes of typical star-forming regions (${\lesssim}0.5\rm\,kpc$; see discussion in \citealt{Anastasopoulou16}). For this reason we restrict our analysis to the 644 galaxies in the host galaxy sample that are closer than $40\Mpc$. This allows for direct comparisons with the works of \citet{Grimm03} and \citet{Mineo12} which adopt similar distance limits.

\subsubsection{Observer bias}
\label{txt:observerbias}

\begin{figure*}
    \centering
    \includegraphics[width=\textwidth]{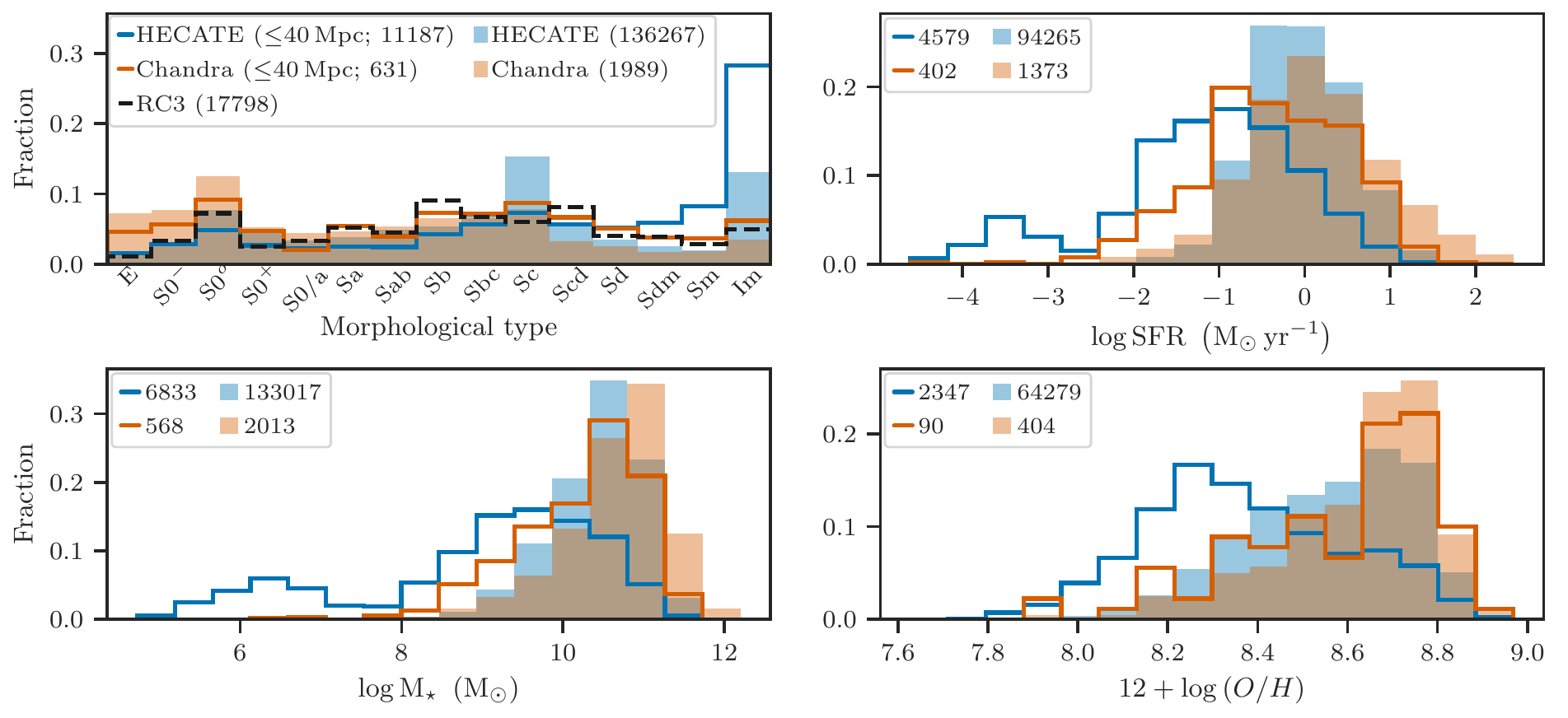}
    \caption{
    Distribution of morphological types (upper left), SFRs (upper right), \mass{} (lower left) and gas-phase metallicities (lower right) for the total volume of the \hec{} (light blue bins) and the $D{<}40\Mpc$ subset (blue steps), as well as, the \hec{}/\CSC{} galaxies (light orange bins) and their $D{<}40\Mpc$ subset (orange steps). For reference, the distribution of morphological types in the {\it RC3} is shown with dashed black steps in the top left panel. The fractions are computed with respect to the sample size of each subset (shown in the legend).
    }
    \label{fig:biases}
\end{figure*}

A limitation of this study is that our ULX sample is not based on a homogeneous, blind survey, but an accumulation of archival data gathered from targeted observations with different selection criteria.
The unknown selection function may lead to observer biases, such as an over-representation of starburst galaxies: SFR is connected to the number of ULXs, as well as, other interesting phenomena (e.g. galaxy mergers) which may have been the focus of \chandra{} observations. To explore any biases or selection effects, \autoref{fig:biases} shows the distributions of the (a) morphological types (see \autoref{tab:rc2} for the morphological classification used in this paper), (b) SFRs, (c) stellar masses, and (d) metallicities for all galaxies with available relevant information in the \hec{}. We compare the parent sample with the subset observed by \chandra{}, in the total volume and the $D{<}40\Mpc$ limited sample.

\begin{table*}
    \caption{Morphological types and corresponding numerical codes (or indices) $T$, as described in the documentation of the Second Reference Catalogue of Bright Galaxies \citep{RC2}. Throughout this paper, we consider as early-type galaxies (ETGs) the elliptical ($T{<}{-}3.5$) and lenticular galaxies ($-3.5{<}T{<}{-}0.5$), and as late-type galaxies (LTGs) the rest. The morphological types of the galaxies are taken from the \hec{}. Throughout the text, different binnings are described as ranges (e.g., Sdm-Im), and measurements with uncertainty less than $1.0$ in $T$ are considered reliable. \label{tab:rc2}
}
\begin{tabular}{l|ccccccccccccccccccc}
    & \multicolumn{3}{l}{Elliptical} & \multicolumn{3}{l}{Lenticular} & \multicolumn{5}{l}{Early spiral} & \multicolumn{4}{l}{Late spiral} & \multicolumn{2}{l}{Irregular (Irr)} \\
    \hline
     Morphological type & cE & E0 & E+ & S0$^-$ & S0$^o$& S0$^+$ & S0/a & Sa & Sab & Sb & Sbc & Sc & Scd & Sd & Sdm & Sm & Im \\
     Numerical code, $T$ & -6 & -5 & -4 & -3 & -2 & -1 & 0 & 1 & 2 & 3 & 4 & 5 & 6 & 7 & 8 & 9 & 10 \\
     \hline
     & \multicolumn{6}{l}{ETGs} & \multicolumn{10}{l}{LTGs}
\end{tabular}
\end{table*}

In terms of morphology, \chandra{} has observed a slightly larger fraction of ETGs galaxies compared to late-type galaxies, a result of observations of nearby clusters which host larger populations of elliptical galaxies. In the $D{<}40\Mpc$ sample, the \hec{} includes a large population of irregular galaxies (mostly satellites of Local Group galaxies), though \chandra{} has observed only a small fraction of them.
For comparison with previous works, in the top left panel of \autoref{fig:biases} we show the distribution of the morphological types in {\it RC3}. The distribution is similar to the one of the host galaxy sample with $D{<}40\Mpc{}$.
We also find that the distributions of \mass{}, SFRs and metallicities of galaxies with \chandra{} observations in the total volume of the \hec{}, are slightly shifted towards larger values than those in the parent sample.

\begin{figure*}
    \centering
    \includegraphics[width=0.99\textwidth]{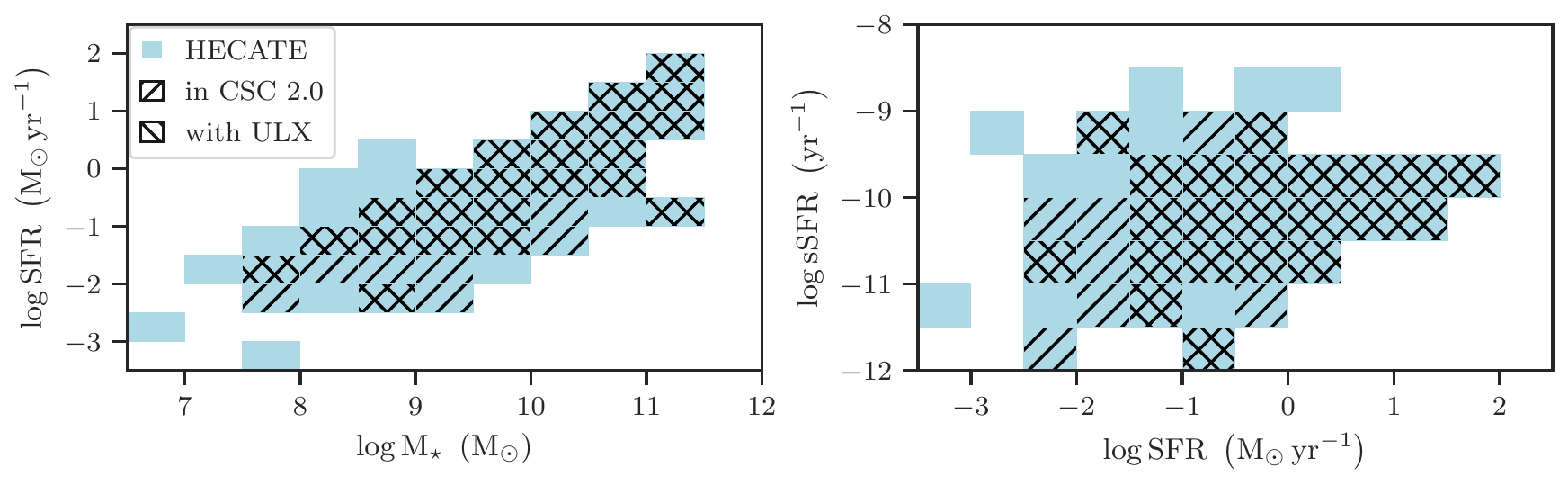}
    \caption{Coverage of the \hec{} non-AGN galaxies with $D{<}40\Mpc$, in the SFR-\mass{} (left) and sSFR-SFR (right) planes. Right-angled hatching indicates galaxies covered by the \CSC{}, while galaxies with left-angled hatching are ULX hosts. It is fairly representative for galaxies of $\mass{>}10^{7.5}\,\rm M_\odot$ and ${\rm SFR}{>}10^{-2.5}\,\rm M_\odot\,yr^{-1}$ and covers the full sSFR range in this region. ULX hosts (back diagonal hatching) cover the same parameter ranges, however, as expected, more sparsely in the low-SFR regime.}
    \label{fig:ulxhostcoverage}
\end{figure*}

\label{txt:representativeness}
These biases combined with the complex selection function of the \hec{} and \chandra{} samples, do not allow us to calculate the volume density of ULXs. Nonetheless, the fact that the X-ray sample covers a wide range of SFRs, \mass{} and metallicities characteristic of the local galaxies, allows us to draw representative scaling relations.
\autoref{fig:ulxhostcoverage} shows the coverage in the SFR-\mass{} and sSFR-SFR planes
for three different samples: the $D{<}40\Mpc$ galaxies in the \hec{}, the subset of those that are included in the \CSC{}, and the subset of the latter hosting ULXs. We note that in this figure we exclude AGN-hosting galaxies to avoid biases in the stellar population parameters (see \S\ref{txt:interlopersAGN}). We find that the host sample covers galaxies down to stellar mass of $10^{7.5}\,\rm M_\odot$ and SFR of $10^{-2.5}\sfrunit{}$, and is uniform in specific SFR (sSFR).

\subsection{Galactocentric distances}
\label{txt:galactrocentric}

The shape of the spatial distributions of ULXs in galaxies of different morphological types can provide valuable information regarding their association to the young or old stellar populations, and globular cluster systems. For this reason we calculate the galactocentric distance as the deprojected distance between the source and the centre of the galaxy, assuming the source resides in the disc of the galaxy. In order to normalise the measured galactocentric distance for the size of the galaxy, we derive the \emph{galactrocentric scale parameter}, $c$ (see \autoref{tbl:sources}), which we define as the ratio of the deprojected galactocentric distance of the source over the semi-major axis of the latter. A full description of the deprojection method and the calculation of $c$ is presented in Appendix~\ref{app:ellipses}.

\subsection{Source luminosities}
\label{txt:luminosities}

In principle, spectral fitting is required for reliable estimates of the source fluxes. Due to the insufficient photon counts for most sources, we use the full-band ($0.5-8.0\rm\, keV$) aperture-corrected net energy flux inferred from the PSF 90\% enclosed count fraction aperture as provided by the \CSC{} (columns (5)-(7) in \autoref{tbl:sources}). In the case of sources with multiple observations, their fluxes are estimated from the \lq{}longest observed segment based on a Bayesian Block analysis of all observations\rq{}\footnote{see \lq{}flux\_aper90\_b\rq{} and \lq{}flux\_aper90\_b\_avg\rq{} in \url{http://cxc.harvard.edu/csc2/columns/fluxes.html} for more details\label{foot:fluxes}.}. We avoid the use of average fluxes (from coadds) since they systematically underestimate the flux of variable sources \citep[e.g.,][]{Zezas07}. Indeed, we find that the above fluxes for our sources are, on average, ${\approx}5\%$ higher than their average\textsuperscript{\ref{foot:fluxes}} fluxes in the \CSC{}.

We convert fluxes to luminosities (columns (12)-(14) in \autoref{tbl:sources}) adopting the distance of the host galaxy in the \hec{} (column (6) in \autoref{tbl:hosts}). The luminosities of 49 sources with significant pileup (column (8) in \autoref{tbl:sources}), are considered as lower limits. However, this does not affect the ULX demographics: their majority (41 sources) are excluded from the ULX demographics as nuclear sources in galaxies hosting an AGN (or without nuclear classification; see \S\ref{txt:interlopersAGN}). Three of them are found in poorly-covered galaxies (excluded from our analysis; see \S\ref{txt:fieldofview}) which are known to not host ULXs (SMC, LMC, Draco Dwarf). The remaining five piled-up sources present luminosities ${>}10^{39}\ergs{}$ and therefore are bona-fide ULX candidates, but their small number does not bias the luminosity distributions presented in this paper, while they are fully accounted for in the demographics.

\subsection{{Foreground/background contamination}}
\label{txt:interlopers}

The main source of contamination in large-area surveys are background (e.g. AGN) and foreground (e.g. stars) objects. Even through we cannot classify individual X-ray sources in our sample as ULXs, AGN, or other classes, using statistical techniques we can remove the effects of these contaminants from our analysis. As a first step we quantify the expected number of foreground/background (f/b) sources in the \CSC{} footprint for each galaxy.

There are two commonly used methods to estimate the surface density of interlopers, based on: (i) blank fields around the target galaxies \citep[e.g.,][]{Wang16}, and (ii) the average $\log{N}$-$\log{S}$ distribution from wide-area and deep surveys \citep[e.g.,][]{Swartz11}.
Since the former method requires around each object the presence of blank areas wide enough to allow the reliable estimation of the interlopers, which is not always the case, we choose the $\log N$-$\log S$ method. We estimate the number of interlopers in the ULX regime in a given galaxy, by rescaling the $\log N$-$\log S$ for the \chandra{}-covered fraction of the $D_{25}$ area of the galaxy, and integrating it down to the flux corresponding to $10^{39}\ergs{}$ for its distance. We use the $\log N$-$\log S$ from the \chandra{} Multiwavelength Project \citep[\textit{ChaMP};][model \lq{}Bc\rq{} in their table~3]{Kim07}. We account for uncertainties in the galaxy distances and the {\it ChaMP} $\log{N}$-$\log{S}$ parameters by Monte Carlo sampling from the corresponding Gaussian error distributions, assuming parameter independence. The expected f/b contamination in each galaxy is given in column (14) of \autoref{tbl:hosts}.

The second step is to estimate the number of bona-fide ULXs in each galaxy given the number of observed sources and the previously calculated background contamination. We model the total number of sources as a mix of ULXs and interlopers, assuming both populations are Poisson distributed. We determine the posterior distribution of the number of ULXs, following the Bayesian method described in \citet{Park06} with the modification that the background \lq{}counts\rq{} in our case are not directly measured but estimated from the $\log{N}$-$\log{S}$. Specifically,
\begin{align*}
\begin{split}
    N_{\rm obs} &= N_{\rm ulx} + N_{\rm f/b} \\
    N_{\rm ulx} &\sim {\rm Pois}(\lambda) \\
    N_{\rm f/b} &\sim {\rm Pois}(\beta),
\end{split}
\end{align*}
where $N_{\rm obs}$ is the observed number of sources in each galaxy: the sum of $N_{\rm ulx}$ ULXs and $N_{\rm f/b}$ interlopers. The latter follow Poisson distributions of means $\lambda$ and $\beta$, respectively, which are independent because the ULX sources in the target galaxy and the foreground/background sources are disconnected populations:
\begin{equation*}
    N_{\rm obs} \sim {\rm Pois}(\lambda + \beta)
    \quad \text{and} \quad
    P(\lambda, \beta) = P(\lambda)P(\beta).
\end{equation*}
To estimate the expected number of ULXs for each galaxy we compute the posterior distribution, marginalised over $\beta$:
\begin{equation*}
    P(\lambda | N_{\rm obs}) = \int\limits_{0}^{\infty}
        P(\lambda, \beta|N_{\rm obs}) {\,\rm d}\beta,
\end{equation*}
where
\begin{equation*}
    P(\lambda, \beta | N_{\rm obs}) \propto
        P(N_{\rm obs}|\lambda, \beta) P(\lambda, \beta)
        =
        P(N_{\rm obs}|\lambda, \beta) P(\lambda) P(\beta).
\end{equation*}
In order to account for uncertainties in the parameters of the $\log N$-$\log S$ distribution, the number of interlopers $\beta$ is not fixed, but allowed to vary. Specifically, the prior for $\beta$ is obtained by evaluating the $\log N$-$\log S$ for varying values of its best-fitting parameters. This is performed by taking $M{=}10000$ samples of the parameters from the corresponding Gaussian distributions (best-fiting values as means, and uncertainties as standard deviations), ultimately giving $M$ samples, $\beta_i$, which represent the distribution of $\beta$. By design, the samples $\beta_i$ have equal probability to be sampled, $P(\beta_i){\propto}1$.
For a uniform prior for $\lambda$, $P(\lambda){\propto}1$, and sufficiently large $M$, the marginalised posterior takes the form
\begin{equation*}
    P(\lambda|N_{\rm obs})
        = P(N_{\rm obs} | \lambda, \beta) P(\beta) P(\lambda)
        \propto \sum\limits_{i=1}^{M} P(N_{\rm obs} | \lambda, \beta_i),
\end{equation*}
where
\begin{equation*}
    P(N_{\rm obs}|\lambda, \beta_i)
    \propto
        \left(\lambda + \beta_i\right)^{N_{\rm obs}} e^{-\lambda-\beta_i}, \quad \lambda \geq 0.
\end{equation*}
From the resulting posterior distribution for each galaxy, we compute the mode and the highest posterior density interval corresponding to $68\%$ probability. The observed number of sources ($N_{\rm obs}$), and the estimate on the number of ULXs ($N_{\rm ulx}$) for each galaxy in our sample are listed in \autoref{tbl:hosts} (columns (13) and (15) respectively).

\begin{figure}
    \centering
    \includegraphics[width=\columnwidth]{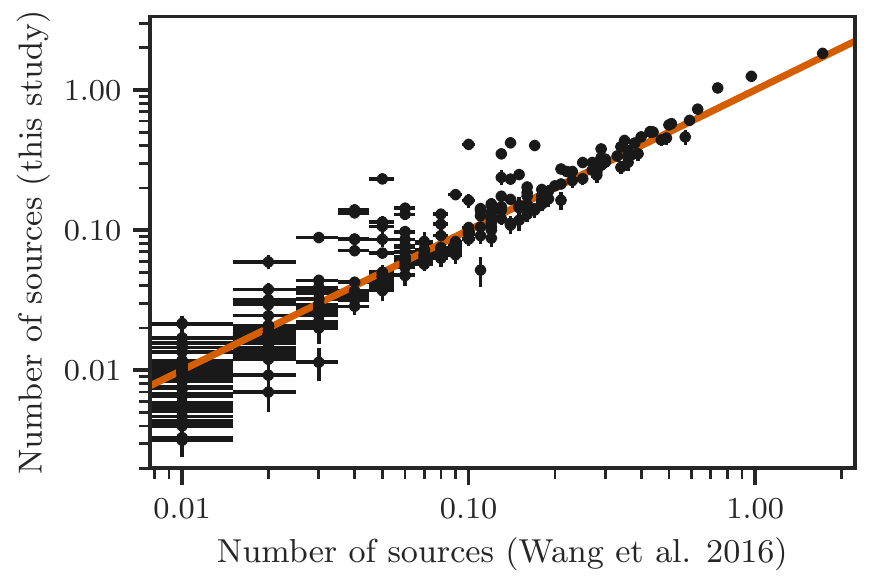}
    \caption{Comparison between the expected foreground/background number of sources from \lq{}blank fields\rq{} approach \citep{Wang16}, and $\log{N}$-$\log{S}$ (this study), down to a luminosity limit of $2{\times}10^{39}\ergs$. The orange line indicates the 1:1 relation. The $x$-error bars are equal to $0.005$ counts because of the $0.01$ precision of the reported values from \citet{Wang16} and the $y$-error bars reflect the uncertainty from the $\log{N}$-$\log{S}$ parameters.}
    \label{fig:compwang}
\end{figure}

To evaluate the accuracy of this method, we compare against the previously published ULX catalogue of \citet{Wang16} which uses the \lq{}blank fields\rq{} approach. We perform this comparison for the 343 galaxies in the sample of \citet{Wang16} that are common with our sample. We adopt the same luminosity cut ($2{\times}10^{39}\,{\rm erg\,s^{-1}}$), distance and area on the sky for each galaxy as \citet{Wang16}. We exclude four very local galaxies from this comparison because the $2\times10^{39}\ergs$ limit corresponds to brighter fluxes than those used to derive the {\it ChaMP} $\log{N}$-$\log{S}$. The results of the comparison of the two methods are shown in \autoref{fig:compwang}. We find that both approaches agree in the total number of interlopers: $33.1{\pm}0.1$ (\lq{}blank fields\rq{}) and $36.4{\pm}0.3$ ($\log N$-$\log S$) interlopers in the galaxies used for this comparison. The results on individual galaxies are in good agreement for the majority of them: $3\sigma$ consistency for 90\% of the objects. A possible explanation for the disagreement of the two methods for the remaining 10\% of the galaxies is the fact that the \lq{}blank fields\rq{} method is based on observations of the individual galaxies, and therefore able to account for the cosmic variance at their location. However, this estimate suffers from incompleteness, due to the degradation of the PSF at the larger off-axis angles from which the background sources are sampled (note that the $\log N$-$\log S$ estimate is on average 10\% higher than the \lq{}blank fields\rq{} estimate).

\subsection{AGN in the host galaxies}
\label{txt:interlopersAGN}

The presence of AGN in the host galaxies can affect our investigation of ULX populations in two ways:
\begin{enumerate}
    \item While AGN typically have $L_{\rm X}{>}10^{42}\,\ergs{}$ \citep[e.g.,][]{Brandt15}, they may exhibit X-ray luminosities as low as $10^{39}$-$10^{40}\ergs{}$ \citep[e.g.,][]{Ho01,Ghosh08,Eracleous02}, and therefore may contaminate the sample of luminous X-ray binaries. We account for this by excluding from the demographics (but not the provided catalogues), the nuclear sources in any galaxies classified as AGN, as well as, in galaxies for which we do not have any information on their nuclear activity. We consider sources as nuclear if they are located in the central $3\asec$ region\footnote{i.e., three times the quadratic sum of the typical positional uncertainty in the \hec{} and the \CSC{} ($1\asec$). The positional uncertainties of the sources are considered negligible since 98\% of the circum-nuclear sources have positional uncertainties ${<}3\arcsec{}$.}. These sources are indicated in column (10) in \autoref{tbl:sources}. Note that this practice unavoidably removes circum-nuclear XRBs \citep[e.g.,][]{Wang16,Gong16}.
    \item The IR component of the AGN emission will overestimate the inferred SFR and \mass{} of the host galaxy, and it will bias the measured metallicity \citep[e.g.,][]{Mullaney11,Delvecchio20}. While the magnitude of this effect can be small in the case of low-luminosity AGN, we take the conservative measure of excluding any AGN hosts (or galaxies with no nuclear activity information) from our scaling relations. However, scaling relations considering all galaxies (regardless of nuclear activity), labelled as \full{} sample to avoid confusion with the non-AGN sample, are presented in Appendix~\ref{app:withagn} and are discussed in the main text when relevant.
\end{enumerate}

\begin{figure}
    \centering
    \includegraphics[width=\columnwidth]{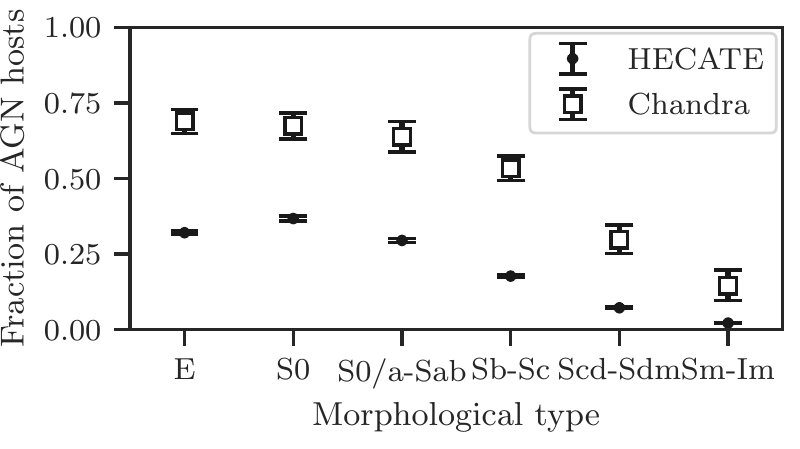}
    \caption{
        The AGN fraction as a function of the morphological type for the \hec{} galaxies (points) and those hosting \chandra{} sources (squares). The error bars correspond to the 68\% CI of the fractions, after accounting for uncertainties on the morphological types, and the Poisson distribution of the number of galaxies.The AGN fraction in \chandra{} targets is higher than in the general population, especially in elliptical galaxies.
        }
    \label{fig:agnfrac}
\end{figure}

To characterise the nuclear activity of the galaxies in our sample, we adopt the classification from the \hec{}, which uses two sources of information to identify AGN:
\begin{enumerate}
    \item
    \citet{Stampoulis19} who classified galaxies as AGN based on their location in 4- or 3-dimensional optical emission-line ratio diagnostic diagrams, using spectroscopic data from the {\it MPA-JHU catalogue}.
    \item
    \citet{She17} who investigated galaxies at $D{<}50\Mpc{}$, observed by \chandra{}: the nuclear classifications are either adopted from the literature or determined using archival optical line-ratio spectral data.
\end{enumerate}
A galaxy is classified as AGN host if it is identified as such in either of the two. These studies provide nuclear activity diagnostics for 539 (84\%) galaxies out of the 644 host galaxies within $40\Mpc$.
\label{txt:agnfraction}
Note, that the exclusion of AGN in the scaling relations, affects the sample of ETGs more strongly since they have higher chance of having been observed due to their nuclear activity, while spiral and irregular galaxies are usually selected for their XRB populations. This is illustrated in \autoref{fig:agnfrac} where we plot the fraction of galaxies with AGN as a function of the morphological type, in the parent galaxy sample and the galaxies with \chandra{} observations.

\section{Results}
\label{txt:results}

In the total volume of the \hec{} we find 23043 X-ray sources, out of which 19260 are characterised as \good{}. In the $D{<}40\Mpc$ sample which is used for the population statistics presented below, there are 16758 \good{} X-ray sources, out of which 793 exceed the ULX limit. Of those 793 sources with $L_{\rm X}{>}\ulxlimit{}$ in the $D{<}40\Mpc$ volume, 164 (21\%) are found close to the centres of galaxies which are classified as AGN hosts, and therefore are not considered as ULX candidates in this study. This leaves a sample of 629 ULX candidates in 309 galaxies, out of which 20\% are expected to be foreground/background contaminants (see \S\ref{txt:lumdist}).

\subsection{Luminosity distribution of X-ray sources}
\label{txt:lumdist}

The luminosity distribution of ULXs is crucial for probing the high-end of the luminosity function (LF) of stellar X-ray sources. The calculation of the LF of the XRBs will be presented in a separate paper. Here, we discuss the distribution of X-ray luminosities above the ULX limit.

\begin{figure}
    \centering
    \includegraphics[width=\columnwidth]{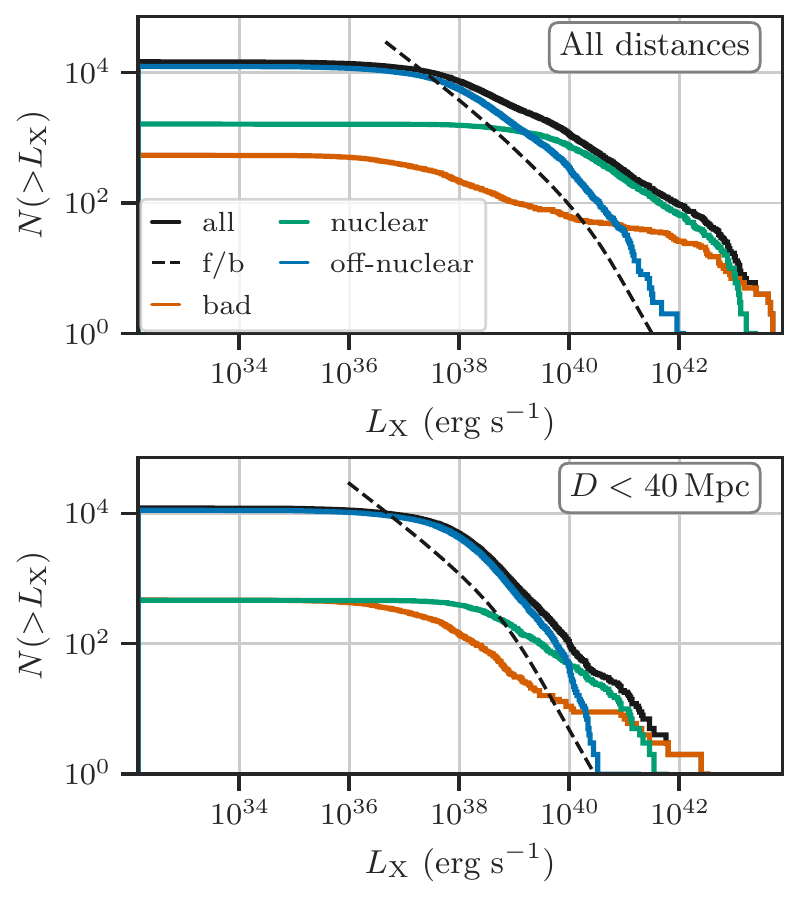}
    \caption{Cumulative number of sources, $N({>}L_{\rm X})$, as a function of X-ray luminosity ($L_{\rm X}$), for various categories: all sources (black), \bad{} sources (orange), off-nuclear \good{} sources (blue), nuclear \good{} sources (green). The expected number of interlopers is shown by the black dashed lines. {\bf Top}: for all sources in our sample. {\bf Bottom}: for sources in galaxies with $D{<}40\Mpc$.}
    \label{fig:lum_dist}
\end{figure}

\autoref{fig:lum_dist} shows the cumulative distribution of the luminosities of the X-ray sources in our sample in all galaxies (top panel) and in those with $D{<}40\Mpc$ (bottom panel). We provide the distribution of all (black), nuclear (green), off-nuclear (blue) and \bad{} sources (orange; see \S\ref{txt:srcquality}). For reference, we also plot the expected distribution of luminosities of interlopers (f/b; see \S\ref{txt:interlopers}). Since we are interested in the contamination of these interlopers within the ULX population, we convert the $\log N$-$\log S$ distribution of the foreground/background sources to the luminosity distribution for the corresponding galaxies using their respective distances. We find that for galaxies within $40\Mpc{}$, the background sources (dashed black line) account for ${\sim}20\%$ of all the sources with $L_{\rm X}{>}10^{39}\ergs$. The contamination dominates the population of the off-nuclear sources at $L_{\rm X}{\gtrsim}10^{40.5}\ergs$.

\label{txt:sensitivity}

The gradual flattening of the luminosity distribution for the total volume (top panel in \autoref{fig:lum_dist}) with decreasing luminosity is a tell-tale sign of incompleteness effects. In the case of the $D{\leq}40\Mpc$ distribution (bottom panel), the flattening in the distribution due to incompleteness occurs at ${\sim}~3{\times}10^{38} \ergs$, below the luminosity limit for ULX candidates in this study. In addition, the typical limiting luminosity of sources detected in the galaxies (using the least luminous source) in the total volume of the \hec{}, ${\sim}4\times10^{39}\ergs$, is above the ULX limit, while in galaxies with $D{<}40\Mpc$, it is ${\sim}1.5{\times}10^{38}\ergs$, well below the luminosity limit for ULX candidates in this study. Therefore, our local sample of ULXs is expected to be complete.

From the top panel of \autoref{fig:lum_dist} we can see that nuclear sources outnumber the off-nuclear sources above $2{\times}10^{39}\ergs$ in the full-volume sample. This is partly the result of the larger distances of galaxies in the full volume survey leading to more significant source confusion: in the dense stellar environment of the galactic cores the sources are blended, ultimately flattening the luminosity distribution. Instead, at the $D{<}40\Mpc$ sample, the source confusion is significantly reduced: the nuclear sources dominate the sample at a higher luminosity ${\sim}10^{40}\ergs$, as it is expected by the population of AGN.

\subsection{Morphology of ULX hosts and spatial distribution of sources}
\label{txt:spatial}

\begin{figure}
    \centering
    \includegraphics[width=\columnwidth]{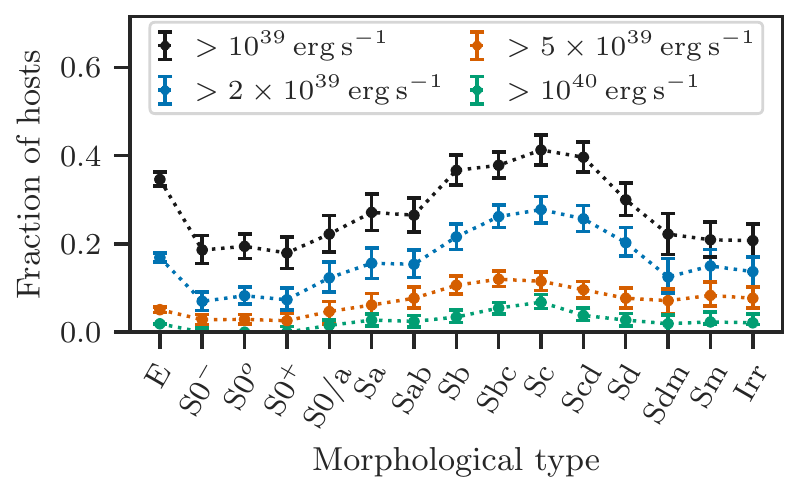}
    \caption{Fraction of galaxies in the \hec{}/\CSC{} hosting at least one off-nuclear source above the ULX limit (black) as a function of the morphological type. Different colours indicate different limits: $2{\times}10^{39}\ergs$ (blue), $5{\times}10^{39}\ergs$ (orange) and $10^{40}\ergs$ (green). The error bars indicate the 68\% CIs, after accounting for the uncertainties on the morphological classifications.}
    \label{fig:fract}
\end{figure}

Early ULX population studies, showed that ULXs preferentially occur in late-type galaxies, with only a small fraction (${\lesssim}20\%$) of elliptical galaxies hosting ULXs \citep[e.g.,][]{Swartz04,Liu06,Swartz11}, in contrast to the recent studies of \citet{Wang16} and \citet{Earnshaw19}. To test this, we quantify the probability for a galaxy to host ULXs as a function of its morphological type. \autoref{fig:fract} shows the fraction of galaxies that host off-nuclear sources above different luminosity thresholds, with respect to all galaxies of the same morphological type with \chandra{} observations. Sources with luminosities above $10^{39}\ergs$ appear to be present in about 30\% of galaxies in all morphological types. There is slightly higher incidence of ULXs (${\sim}40\%$) in elliptical galaxies and Sb-Scd spiral galaxies, while lenticular and irregular galaxies are less likely to host ULXs (${\sim}20\%$). However, galaxies containing sources with $L_{\rm X}{>}5{\times}10^{39}\ergs$ are typically of late type.

The spatial distribution of X-ray sources can provide insights into their nature. We would expect the surface density of ULXs to follow the distribution of starlight in the host galaxies. However, \citet{Swartz11} and \citet{Wang16} find a flattening of the surface density of ULXs in spiral galaxies at large galactocentric distance, in contrast to their exponential surface brightness profile. In addition, \citet{Wang16} observe an excess of ULXs at large galactocentric distances in elliptical galaxies.

\begin{figure*}
    \centering
    \includegraphics[width=\textwidth]{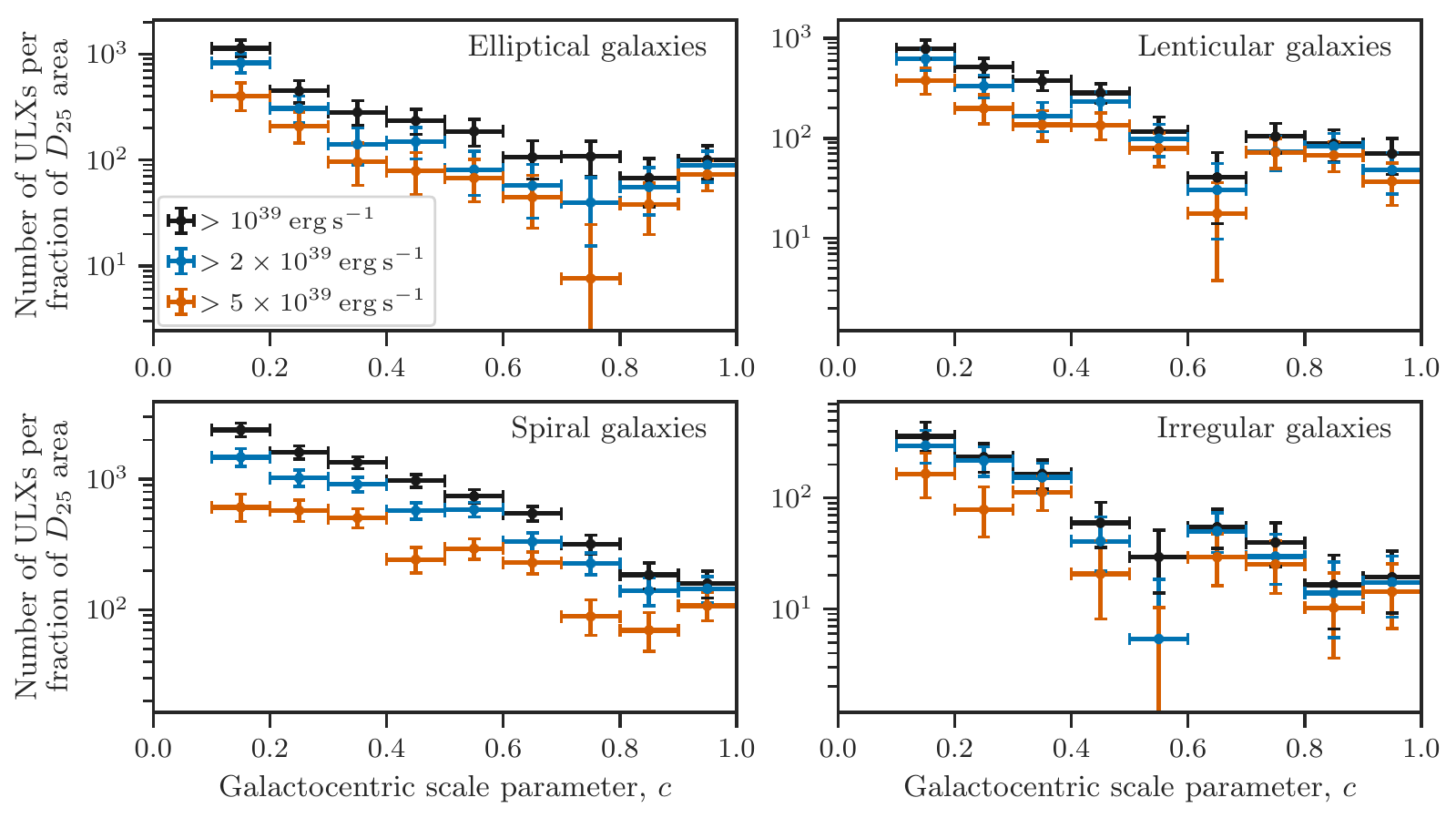}
    \caption{The density of ULXs per fraction of $D_{25}$ area enclosed in annuli with galactocentric distances $c-{\rm d}c$ to $c+{\rm d}c$, as function of $c$, for different luminosity thresholds (colours), and morphological classes of galaxies. The estimates are corrected for f/b contamination. The dips observed in lenticular (at $c{\sim}0.65$) and irregular galaxies (at $c{\sim}0.55$) are statistical fluctuations due to the small numbers of sources; they were found to be sensitive to the binning scheme.}
    \label{fig:galcen_t_ulx}
\end{figure*}

In order to test these observations, we quantify the spatial distribution of ULXs, by computing their surface density on the basis of their galactocentric distances, $c$, for off-nuclear sources with luminosities above $1$, $2$ and $5{\times} 10^{39}\ergs$. Using the method described in \S\ref{txt:interlopers}, we correct for the expected f/b contamination. We perform this exercise for galaxies of different morphological types and distances up to $40\Mpc$. The distributions are shown in \autoref{fig:galcen_t_ulx}.
We do not report the number of ULXs at $c{<}0.1$ since it is biased by the exclusion of nuclear sources\footnote{At $40\Mpc$ the typical semi-major axis of galaxies in our sample is $30\asec$, while in this study we consider nuclear regions of $3\asec$.}.

We find that the surface density of ULXs follows the expected exponential trend in spiral galaxies, in contrast to \citet{Swartz04} and \citet{Wang16}. This disagreement may be caused by foreground/background contamination since it is not accounted for in the spatial distribution analysis of \citet{Swartz04} and \citet{Wang16}. The number of interlopers scales with area and therefore adds a constant in the surface density profile, effectively flattening the distributions. In our sample, the surface density of ULXs in elliptical galaxies flattens at large galactocentric distances as observed by \citet{Wang16}.

\subsection{Rate of ULXs}
\label{txt:rateulx}

To quantify the number of ULXs per galaxy as a function of their luminosity for various morphological types,  we consider five luminosity thresholds: 1, 2, 3, 5 and 10${\times}10^{39}\ergs$. We compute the background-corrected number of ULXs above each luminosity threshold, and its 68\% confidence intervals, by accumulating the number of observed sources and expected interlopers, and applying the method described in \S\ref{txt:interlopers}. The calculation is performed for each morphological class, as well as, the total galaxy sample. Galaxies with uncertain morphological classification are excluded from this analysis (uncertainty in numerical morphological code ${>}1$; see \autoref{tab:rc2}). By considering non-AGN host galaxies, we also calculate the number of ULXs per total SFR for LTGs, or per total \mass{} for ETGs. In addition, we also report the number of ULXs per total \mass{} for all ETGs (\lq{}full\rq{}). The results are presented in \autoref{tbl:ulxrates}.

\newcommand{\grey}[1]{{\color{gray!50!black}\ #1}}
\begin{table*}
    \caption{Background-corrected number of ULXs above certain luminosity limits (first line) for different host morphological classes (first column). Square brackets indicate the total number of observed sources followed by the expected number of interlopers. The confidence intervals (68\%) account for Poisson uncertainties in both the number of observed sources and the number of interlopers.
    \textbf{(A)}: Number of ULXs divided by the number of host galaxies (second column). The last row (total) refers to all morphological classes.
    \textbf{(B)}: The number of ULXs per $\sfrunit$ of SFR (third column). The last row (LTGs) refers to all late-type galaxies.
    \textbf{(C)}: The number of ULXs per $10^{12}\massunit$ stellar mass (third column). The last row (ETGs) refers to all early-type galaxies.
    \textbf{(D)}: As (C), but now considering the full sample of ETGs (not excluding AGN hosts).
    }
\label{tbl:ulxrates}
\flushleft
\begin{tabular}{lrrlllll}
\multicolumn{8}{c}{(A) \it Number of ULXs per galaxy}\\\hline
Morph. type & $N_{\rm gal}$ &
                & ${>}10^{39}\ergs$
                & ${>}2{\times}10^{39}\ergs$
                & ${>}3{\times}10^{39}\ergs$
                & ${>}5{\times}10^{39}\ergs$
                & ${>}10^{40}\ergs$
                \\\hline
E               &   101 & & $0.68^{+0.10}_{-0.10}$ \grey{[105|36.6]} &
                            $0.32^{+0.07}_{-0.06}$ \grey{[47|15.0]} &
                            $0.23^{+0.06}_{-0.05}$ \grey{[32|8.4]} &
                            $0.16^{+0.05}_{-0.04}$ \grey{[20|3.9]} &
                            $0.03^{+0.02}_{-0.02}$ \grey{[4|1.4]}
                            \\[2pt]
S0              &  99   & & $0.28^{+0.07}_{-0.07}$ \grey{[47|18.8]} &
                            $0.11^{+0.05}_{-0.04}$ \grey{[18|7.6]} &
                            $0.09^{+0.04}_{-0.03}$ \grey{[13|4.2]} &
                            $0.05^{+0.03}_{-0.02}$ \grey{[7|2.0]} &
                            $0.00^{+0.01}_{-0.00}$ \grey{[0|0.7]}
                            \\[2pt]
S0/a-Sb         &  89   & & $1.02^{+0.12}_{-0.12}$ \grey{[115|24.3]} &
                            $0.48^{+0.08}_{-0.08}$ \grey{[52|9.5]} &
                            $0.38^{+0.07}_{-0.07}$ \grey{[39|5.2]} &
                            $0.25^{+0.06}_{-0.05}$ \grey{[25|2.5]} &
                            $0.09^{+0.04}_{-0.03}$ \grey{[9|0.9]}
                            \\[2pt]
Sbc-Sd          &  166  & & $1.38^{+0.10}_{-0.09}$ \grey{[258|28.5]} &
                            $0.76^{+0.07}_{-0.07}$ \grey{[138|11.7]} &
                            $0.48^{+0.06}_{-0.05}$ \grey{[87|6.6]} &
                            $0.28^{+0.04}_{-0.04}$ \grey{[50|3.1]} &
                            $0.12^{+0.03}_{-0.03}$ \grey{[21|1.1]}
                            \\[2pt]
Sdm-Im          &   37  & & $0.48^{+0.13}_{-0.11}$ \grey{[20|2.1]} &
                            $0.30^{+0.10}_{-0.08}$ \grey{[12|0.8]} &
                            $0.26^{+0.10}_{-0.08}$ \grey{[10|0.4]} &
                            $0.18^{+0.08}_{-0.06}$ \grey{[7|0.2]} &
                            $0.03^{+0.04}_{-0.02}$ \grey{[1|0.1]}
                            \\[2pt]
{\bf Total}     &  492  & & $0.88^{+0.05}_{-0.05}$ \grey{[545|110.3]} &
                            $0.45^{+0.03}_{-0.03}$ \grey{[267|44.5]} &
                            $0.32^{+0.03}_{-0.03}$ \grey{[181|24.9]} &
                            $0.20^{+0.02}_{-0.02}$ \grey{[109|11.7]} &
                            $0.06^{+0.01}_{-0.01}$ \grey{[35|4.2]}
\\\hline\\
\multicolumn{8}{c}{(B) \it Number of ULXs per $\sfrunit$ SFR in non-AGN, late-type galaxies}
\\\hline
%
%
Morph. type          & $N_{\rm gal}$ & SFR
                & ${>}10^{39}\ergs$
                & ${>}2{\times}10^{39}\ergs$
                & ${>}3{\times}10^{39}\ergs$
                & ${>}5{\times}10^{39}\ergs$
                & ${>}10^{40}\ergs$
                \\\hline
S0/a-Sb         &  23  & 86.5  &
                            $0.29^{+0.07}_{-0.06}$ \grey{[30|4.6]} &
                            $0.17^{+0.05}_{-0.04}$ \grey{[16|1.7]} &
                            $0.14^{+0.05}_{-0.04}$ \grey{[13|1.0]} &
                            $0.06^{+0.03}_{-0.02}$ \grey{[6|0.4]} &
                            $0.02^{+0.02}_{-0.01}$ \grey{[2|0.2]}
                            \\[2pt]
Sbc-Sd          &  76  & 97.9  &
                            $0.78^{+0.10}_{-0.09}$ \grey{[87|10.6]} &
                            $0.40^{+0.07}_{-0.06}$ \grey{[43|4.1]} &
                            $0.22^{+0.05}_{-0.05}$ \grey{[24|2.3]} &
                            $0.12^{+0.04}_{-0.03}$ \grey{[13|1.1]} &
                            $0.07^{+0.03}_{-0.02}$ \grey{[7|0.4]}
                            \\[2pt]
Sdm-Im          &  20  & 2.3  &
                            $2.39^{+1.21}_{-0.91}$ \grey{[6|0.4]} &
                            $1.22^{+0.91}_{-0.61}$ \grey{[3|0.1]} &
                            $1.25^{+0.91}_{-0.61}$ \grey{[3|0.1]} &
                            $0.84^{+0.79}_{-0.48}$ \grey{[2|0.0]} &
                            $0.00^{+0.48}_{-0.00}$ \grey{[0|0.0]}
                            \\[2pt]
{\bf LTGs}     &  119  & 186.7  &
                            $0.58^{+0.06}_{-0.08}$ \grey{[123|15.5]} &
                            $0.30^{+0.04}_{-0.04}$ \grey{[62|6.0]} &
                            $0.20^{+0.04}_{-0.03}$ \grey{[40|3.3]} &
                            $0.10^{+0.03}_{-0.02}$ \grey{[21|1.6]} &
                            $0.05^{+0.02}_{-0.01}$ \grey{[9|0.6]}
\\\hline\\
\multicolumn{8}{c}{(C) \it Number of ULXs per $10^{12}\massunit$ stellar mass in non-AGN, early-type galaxies}
\\\hline
%
%
Morph. type          & $N_{\rm gal}$ & $\mass{}$
                & ${>}10^{39}\ergs$
                & ${>}2{\times}10^{39}\ergs$
                & ${>}3{\times}10^{39}\ergs$
                & ${>}5{\times}10^{39}\ergs$
                & ${>}10^{40}\ergs$
                \\\hline
E          &   22 & 0.82  &
                            $23.7^{+6.5}_{-5.6}$ \grey{[25|5.4]} &
                            $19.2^{+5.5}_{-4.7}$ \grey{[18|2.2]} &
                            $16.7^{+5.1}_{-4.3}$ \grey{[15|1.2]} &
                            $13.9^{+4.6}_{-3.8}$ \grey{[12|0.6]} &
                            $\phantom{0}1.0^{+1.7}_{-0.9}$ \grey{[1|0.2]} 
                            \\[2pt]
S0              &   28 & 0.73  &
                            $10.1^{+4.8}_{-3.9}$ \grey{[10|2.6]} &
                            $\phantom{0}8.1^{+4.1}_{-3.2}$ \grey{[7|1.0]} &
                            $\phantom{0}6.0^{+3.5}_{-2.6}$ \grey{[5|0.6]} &
                            $\phantom{0}3.7^{+2.9}_{-2.0}$ \grey{[3|0.3]} &
                            $\phantom{0}0.0^{+1.5}_{-0.0}$ \grey{[0|0.1]}
                            \\[2pt]
{\bf ETGs}     &   50 & 1.56  &
                            $17.3^{+4.0}_{-3.6}$ \grey{[35|8.0]} &
                            $13.9^{+3.4}_{-3.0}$ \grey{[25|3.2]} &
                            $11.7^{+3.1}_{-2.7}$ \grey{[20|1.8]} &
                            $\phantom{0}9.1^{+2.7}_{-2.3}$ \grey{[15|0.8]} &
                            $\phantom{0}0.5^{+0.9}_{-0.5}$ \grey{[1|0.3]}
\\\hline\\
\multicolumn{8}{c}{(D) \it Number of ULXs per $10^{12}\massunit$ stellar mass in all early-type galaxies}
\\\hline
%
%
Morph. type          & $N_{\rm gal}$ & $\mass{}$
                & ${>}10^{39}\ergs$
                & ${>}2{\times}10^{39}\ergs$
                & ${>}3{\times}10^{39}\ergs$
                & ${>}5{\times}10^{39}\ergs$
                & ${>}10^{40}\ergs$
                \\\hline
E (full)          &   96 & 7.8  &
                            $8.72^{+1.34}_{-1.26}$ \grey{[105|36.6]} &
                            $4.09^{+0.91}_{-0.83}$ \grey{[47|15.0]} &
                            $3.01^{+0.76}_{-0.68}$ \grey{[32|8.4]} &
                            $2.05^{+0.61}_{-0.53}$ \grey{[20|3.9]} &
                            $0.33^{+0.30}_{-0.21}$ \grey{[4|1.4]} 
                            \\[2pt]
S0 (full)             &   98 & 5.2  &
                            $5.37^{+1.36}_{-1.24}$ \grey{[47|18.8]} &
                            $1.98^{+0.87}_{-0.74}$ \grey{[18|7.6]} &
                            $1.67^{+0.75}_{-0.62}$ \grey{[13|4.2]} &
                            $0.95^{+0.57}_{-0.44}$ \grey{[7|2.0]} &
                            $0.00^{+0.22}_{-0.00}$ \grey{[0|0.7]}
                            \\[2pt]
{\bf ETGs} (full)     &   194 & 13.1  &
                            $7.38^{+0.96}_{-0.91}$ \grey{[152|55.4]} &
                            $3.24^{+0.64}_{-0.59}$  \grey{[65|22.5]} &
                            $2.47^{+0.54}_{-0.49}$ \grey{[45|5.9]} &
                            $1.61^{+0.42}_{-0.37}$  \grey{[27|5.9]} &
                            $0.14^{+0.16}_{-0.12}$  \grey{[4|2.1]}
\\\hline
\end{tabular}
\end{table*}

We find that the number of ULXs per galaxy for the total population is $0.88{\pm}0.05$ for $L_{\rm X}{>}10^{39}\ergs$ and $0.20{\pm}{0.02}$ for $L_{\rm X}{>}5{\times}10^{39}\ergs$.
Comparable frequencies are found in early spirals (S0/a-Sb), while in late spirals (Sbc-Sd) they are ${\sim}1.5$ times higher. In elliptical galaxies (E) the ULX frequency per galaxy is slightly lower than that of the total population. Lenticular (S0) galaxies present the lowest frequencies in all luminosity limits, with irregular galaxies following.

The number of ULXs per SFR in irregular galaxies (Sdm-Im) is higher than in spirals, in contrast to their small numbers per galaxy. Early spirals (S0/a-Sb) exhibit the lowest numbers of ULXs per SFR.

Additionally, the \full{} sample of ETGs presents lower specific ULX frequencies than the non-AGN sample by a factor of ${\sim}2$. Interestingly, we find that the number of ULXs per \mass{} is higher in elliptical (E) than in lenticular galaxies (S0) by a factor of ${\sim}2$, a result also observed by \citet{Wang16}. However, this trend disappears when considering the \full{} ETG sample: the specific ULX frequencies are consistent within the uncertainties.

\label{txt:earlylate}
\begin{table}
    \caption{Fitting results for the scaling factor $a$ (\autoref{eq:modela}) for all LTGs and different morphological classes, and the scaling factor $b$ (\autoref{eq:modelb}) for all ETGs, elliptical (E) and lenticular (S0) galaxies. These results are plotted in \autoref{fig:earlylatefits_nAGN}.
}
\begin{tabular}{lrrrr@{\hskip 2pt}l}
    \hline
    Morphology & $N_{\rm gal}$ & $N_{\rm src}$ & $N_{\rm f/b}$ & \multicolumn{2}{c}{$a$ (\aunits)} \\\hline
    LTGs     & 119 & 123 & 17.0 & \hspace{0.7cm} $0.51$ & $^{+0.06}_{-0.06}$ \\[2pt]
    S0/a-Sab &  11 &  18 &  2.9 & $0.23$ & $^{+0.07}_{-0.07}$ \\[2pt]
    Sb       &  12 &  12 &  2.2 & $0.34$ & $^{+0.14}_{-0.12}$ \\[2pt]
    Sbc      &  14 &  26 &  4.8 & $0.44$ & $^{+0.12}_{-0.11}$ \\[2pt]
    Sc       &  17 &  19 &  2.6 & $0.51$ & $^{+0.15}_{-0.13}$ \\[2pt]
    Scd      &  22 &  23 &  2.4 & $1.54$ & $^{+0.39}_{-0.34}$ \\[2pt]
    Sd-Sdm   &  23 &  19 &  1.7 & $2.29$ & $^{+0.67}_{-0.57}$ \\[2pt]
    Sm-Im    &  20 &   6 &  0.5 & $2.16$ & $^{+1.19}_{-0.90}$ \\\hline
    Morphology & $N_{\rm gal}$ & $N_{\rm src}$ & $N_{\rm f/b}$ & \multicolumn{2}{c}{$b$ ($\bunits$)} \\\hline
    ETGs        & 50 & 35 & 8.3 & $15.1$ & $^{+3.9}_{-3.6}$ \\[2pt]
    E           & 22 & 25 & 5.5 & $21.9$ & $^{+6.4}_{-5.7}$ \\[2pt]
    S0          & 28 & 10 & 2.8 &  $8.7$ & $^{+4.7}_{-3.8}$ \\[2pt]
    ETGs (full) &195 &152 &57.6 &  $6.3$ & $^{+1.0}_{-0.9}$ \\[2pt]
    E    (full) & 96 &104 &37.5 &  $7.5$ & $^{+1.3}_{-1.3}$ \\[2pt]
    S0   (full) & 99 & 47 &20.0 &  $4.8$ & $^{+1.4}_{-1.2}$ \\\hline
\end{tabular}
\label{tbl:earlylate}
\end{table}

\begin{figure}
    \centering
    \includegraphics[width=\columnwidth]{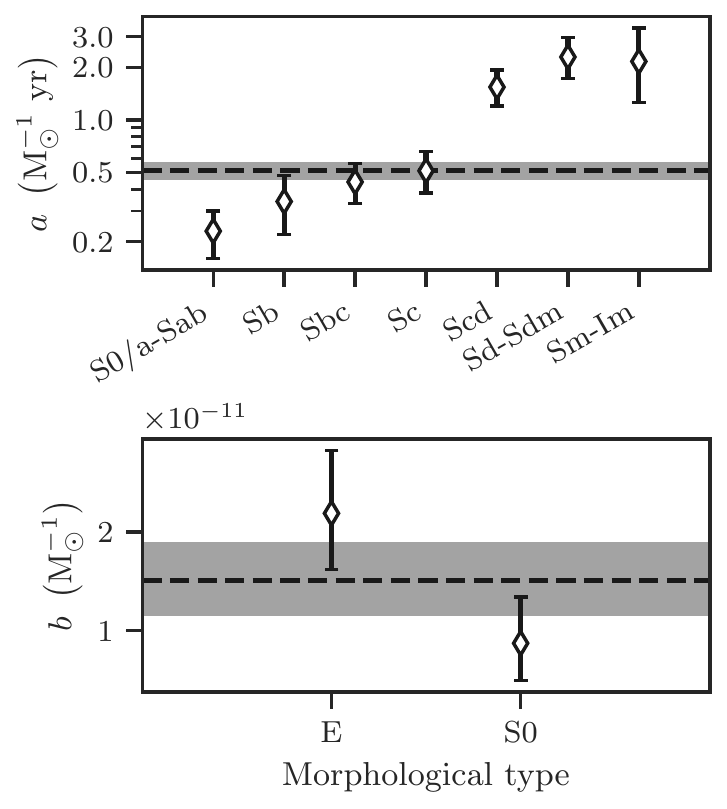}
    \caption{Fitting results for the SFR and \mass{} scaling factors for different morphological classes.
    \textbf{Top}: the scaling parameter $a$ (\autoref{eq:modela}) for late-type galaxies (line) and 68\% CI (grey band), and for various late-type morphological classes (black error bars).
    \textbf{Bottom}: same as the top panel, but for the scaling parameter $b$ (\autoref{eq:modelb}) for early-type (line and band), elliptical and lenticular galaxies (error bars).}
    \label{fig:earlylatefits_nAGN}
\end{figure}

In addition to the above analysis, we perform fits of the number of ULXs against the \mass{} or SFR of the galaxies. For $N_{\rm obs}$ sources with $L_{\rm X}{>}10^{39}\ergs{}$ and $N_{\rm f/b}$ expected number of interlopers from the \textit{ChaMP} $\log N$-$\log S$, we fit the model
\begin{equation}
    N_{\rm obs} = {\rm Pois}\left(b \times \mass{} + N_{\rm f/b}\right)
    \label{eq:modelb}
\end{equation}
for all ETGs with robust morphological classifications (see caption of \autoref{tab:rc2}), and the subdivisions of elliptical and lenticular galaxies, using the Maximum Likelihood method.
In a similar fashion, we fit the model
\begin{equation}
    N_{\rm obs} = {\rm Pois}\left(a \times \text{SFR} + N_{\rm f/b}\right)
    \label{eq:modela}
\end{equation}
for the LTGs and five sub-populations: S0/a-Sab, Sb, Sbc, Sc, Scd, Sd-Sdm and Sm-Im\footnote{The selection of the ranges of morphological types ensured that at least ten galaxies were contributing to the statistical estimates.}. The results are listed in \autoref{tbl:earlylate}.

For the scaling with \mass{} in ETGs we find $b{=}15.1^{+3.9}_{-3.6}\bunits{}$, while in elliptical galaxies it is significantly higher ($21.9^{+6.4}_{-5.7}\bunits{}$) than in lenticular galaxies ($8.7^{+4.7}_{-3.8}\bunits{}$). However, in the full ETG sample (i.e. including AGN hosts), the specific ULX frequencies are lower than those in the non-AGN ETGs ($6.3^{+1.0}_{-0.9}\bunits$). See \S\ref{txt:disearly} for an explanation of this difference.

In LTGs, we find that the scaling with SFR, $a$, is $0.51{\pm}0.06\aunits{}$ (horizontal line in the top panel of \autoref{fig:earlylatefits_nAGN}) and that it monotonically increases with morphological type: from $0.23\aunits{}$ (S0/a-Sab) to $2.16\aunits{}$ (Sm-Im).

\subsection{SFR and stellar mass scaling in late-type galaxies}
\label{txt:lateboth}

In order to account for the contribution of ULXs associated with LMXBs (e.g., GRS1915+105-type systems; \citealt{Greiner01}), we perform a joint fit of the number of ULXs in LTGs with respect to both their SFR and \mass{}.

\begin{figure}
    \centering
    \includegraphics[width=\columnwidth]{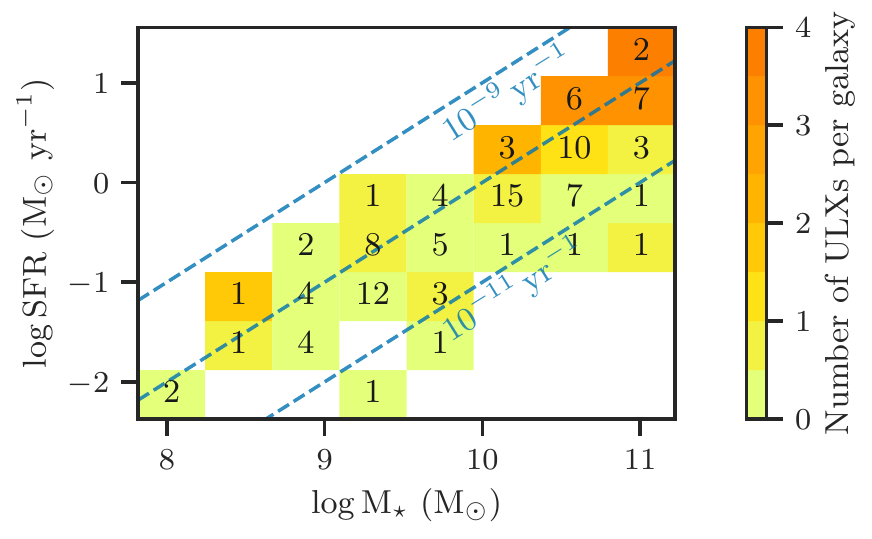}
    \caption{
        The mean, background-corrected number of ULXs ($N_{\rm ulx}$) per galaxy (color scale) as a function of SFR and \mass{}, in non-AGN LTGs. The diagonal dashed lines correspond to indicative specific SFRs, while the numbers in boxes denote the number of galaxies in each bin. We see a trend for more ULXs in galaxies with high SFR, \mass{}, and sSFR.
    }
    \label{fig:msfrplane}
\end{figure}

The correlation can be visualised by binning the galaxies by SFR and \mass{} and computing the average number of ULXs per galaxy, for each bin, after removing the f/b contamination (see \S\ref{txt:interlopers}). We plot the result in \autoref{fig:msfrplane}, where we see a trend for galaxies with higher SFR and \mass{} to host larger numbers of ULXs. This trend becomes stronger in regions of high sSFR (indicated by the diagonal lines).

\begin{figure}
\captionof{table}{
    Mode and 68\% Highest Posterior (marginalised) Density Intervals of the scaling parameters $\alpha$ and $\beta$ (see \autoref{eq:modelab}) for all LTGs and their \lq{}early\rq{} and \lq{}late\rq{} subdivisions (see \autoref{tab:rc2}). For each fit we report the number of galaxies ($N_{\rm gal}$), sources ($N_{\rm src}$) and interlopers ($N_{\rm f/b}$). See below for the joint distributions (\autoref{fig:2dfits_nAGN}).
    \label{tbl:2dfits}
}
\begin{tabular}{lrrrcc}
    \hline
    Sample & $N_{\rm gal}$ & $N_{\rm src}$ & $N_{\rm f/b}$ & $\alpha\ \rm (M_\odot\,yr^{-1})^{-1}$ & $\beta\ (10^{12}\,\rm M_\odot)^{-1}$ \\\hline 
    LTGs    & 106 & 117 & 15.8 & $0.45^{+0.06}_{-0.09}$ & $\phantom{{<}0}3.3^{+3.8}_{-3.2}$ \\[2pt]
    S0/a-Sbc & 37  & 56  &  9.9 & $0.16^{+0.08}_{-0.08}$ & $\phantom{<}11.2^{+5.2}_{-5.6}$ \\[2pt]
    Sc-Im    & 69  & 61  &  5.9 & $0.98^{+0.11}_{-0.20}$ & $\phantom{0}{<}6.6\phantom{^{+0.0}_{-0.0}}$ \\\hline
\end{tabular}

\includegraphics[width=\columnwidth]{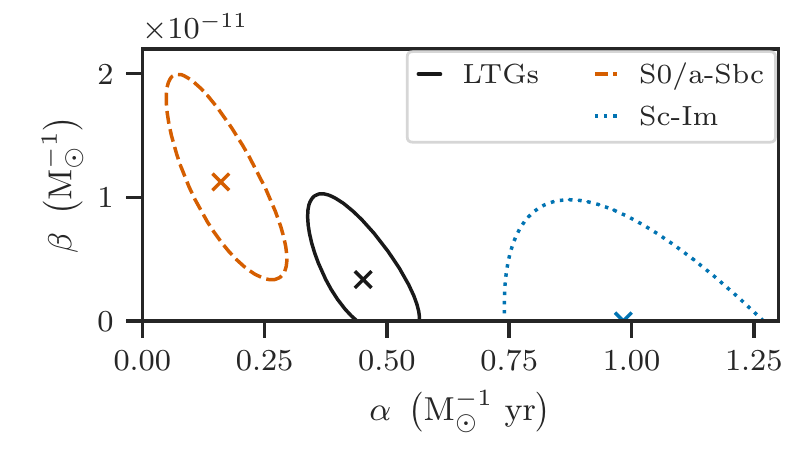}
\captionof{figure}{
    The best-fitting values (X symbols) and, 68\% confidence regions (lines) of the scaling parameters  $a$ and $b$ of \autoref{eq:modelab} for all LTGs (solid), and the \lq{}early\rq{} (S0/a-Sbc; orange) and \lq{}late\rq{} (Sc-Im; blue) spiral galaxies. See \autoref{tbl:2dfits} for marginalised results.
    \label{fig:2dfits_nAGN}
}
\end{figure}

While SFR and \mass{} are known to be correlated in star-forming galaxies \citep[e.g.,][]{Rodighiero11,Speagle14,Maragkoudakis17}, the SFR is expected to be the primary parameter correlated with the population of ULXs. To study the dependence of $N_{\rm ulx}$ on both parameters, we fit the model
\begin{equation}
    N_{\rm obs} \sim {\rm Pois}\left( \alpha \times {\rm SFR} + \beta \times \mass{} + N_{\rm f/b}\right)
    \label{eq:modelab}
\end{equation}
where $N_{\rm obs}$ is the total number of observed sources with $L_{\rm X}{>}10^{39} \ergs$, $\alpha$ and $\beta$ are the scaling factors that will be fitted, and $N_{\rm f/b}$ is the expected number of interlopers (computed in \S\ref{txt:interlopers}).
The model is applied to all LTGs with robust morphological classifications (see Tables~\ref{tbl:hosts}, \ref{tab:rc2}) and the two sub-populations of early spirals (S0/a-Sbc), and late spirals / irregular galaxies (Sc-Im). The results are listed in \autoref{tbl:2dfits}, while the joint posterior distributions of $\alpha$ and $\beta$ are shown in \autoref{fig:2dfits_nAGN}.

The best-fitting value of the SFR scaling factor for LTGs is $\alpha{=}0.45^{+0.06}_{-0.09}\aunits$ while the \mass{} scaling factor is $\beta{=}3.3^{+3.8}_{-3.2}\bunits$. For the early-type spirals we find lower $\alpha{=}0.16{\pm}0.08\aunits$ and higher $\beta{=}11.2^{+5.2}_{-5.6}\bunits$, while for the late-type spirals the situation is inverted, i.e. $\alpha{=}0.98^{+0.11}_{-0.20}\aunits$ and $\beta$ is consistent with zero (${<}6.6\bunits{}$).

\section{Discussion}
\label{txt:discussion}

\subsection{Comparison with other ULX surveys}
\label{txt:discomp}

Our estimate for the scaling of the number of ULXs with SFR ($0.51\pm0.06 \aunits{}$; \autoref{tbl:earlylate}) is four times lower than that estimated in \citeauthor{Swartz11} (\citeyear{Swartz11}; 2\aunits{}). This is the result of differences in: (i) the selection of host galaxy sample, and (ii) the method used in the calculation of the X-ray fluxes. When we account for these differences, we find consistent results as discussed in detail in Appendix~\S\ref{app:swartz}.

Furthermore, we would expect our results to agree with those of \citet{Wang16}, since they also use \chandra{} observations for a similarly large sample of host galaxies (343 galaxies) to study the ULX content in nearby galaxies. However, \citet{Wang16} consider ULXs at twice our luminosity threshold (i.e., $2\times10^{39}$) and at larger separations from the galaxy centres ($2{\times}D_{25}$ area instead of $D_{25}$). After accounting for these differences, and a small offset between the computed X-ray fluxes resulting from different methods, we find similar frequency of ULXs in all galaxies, and separately for their different morphological classes (\autoref{tbl:ulxrates}). See Appendix~\S\ref{app:wang} for details of this comparison.

\citet{Earnshaw19}, using a sample of 248 galaxies with sensitivity limit below the ULX limit in their X-ray samples, found that one out of three galaxies host at least one ULX, with spiral galaxies having a slightly higher fraction (${\sim}40\%$) than elliptical galaxies (${\sim}30\%$). This is in agreement with our results (see \autoref{fig:fract}): the fraction of ULX hosts in galaxies of different morphological types is between 20\% and 40\%, with the peak at Sc galaxies, and a fraction of ${\sim}35\%$ in elliptical galaxies.

\subsection{Dependence of number of ULXs on SFR and stellar mass in star-forming galaxies}
\label{txt:dislate}

\begin{figure}
    \centering
    \includegraphics[width=\columnwidth]{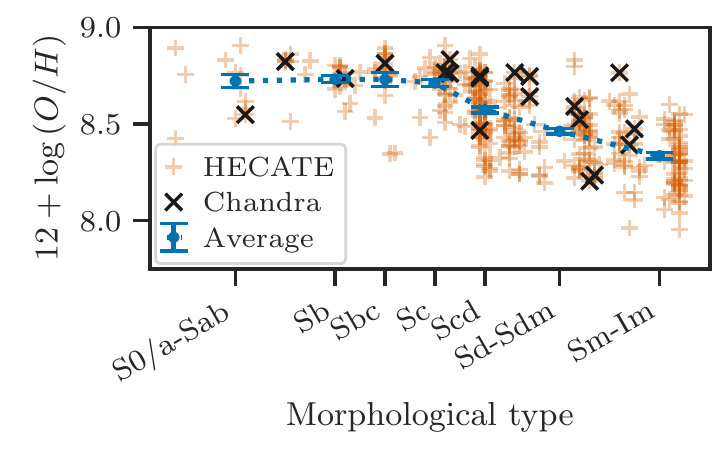}
    \caption{
        The metallicity, $12+\log\left(O/H\right)$, as a function of the morphological type in the host (\lq{}Chandra\rq{}; black \lq{}x\rq{} markers) and parent (\lq{}HECATE\rq{}; orange plus markers) galaxy samples. Note, that only non-AGN galaxies with robust morphological classification and $D{<}40\Mpc{}$ are shown. For the morphological types that correspond to the results of the scaling of ULXs with SFR, $a$, in \autoref{tbl:earlylate}, we show the mean metallicity and its standard error (blue errorbars). Scd galaxies and later, present the lower metallicities, partly explaining the result that in the same galaxies $a$ is significantly higher than the average in LTGs.
        \label{fig:metalvst}
    }
    \label{fig:my_label}
\end{figure}

In \S\ref{txt:earlylate}, we find the number of ULXs in LTGs to be $0.51{\pm}0.06$ per \sfrunit{} (see \autoref{tbl:earlylate}), consistent with the expectation from the \citet{Mineo12} HMXB-LF, of $0.56\aunits{}$. We observe a dependence of the scaling factor per SFR (parameter $a$ in \autoref{eq:modela} and \autoref{fig:earlylatefits_nAGN}) on the morphological type; it monotonically increases from $0.23{\pm}0.07$ to $2.16^{+1.19}_{-0.90}\aunits$ from S0/a-Sab to Sm-Im galaxies (\autoref{tbl:earlylate}). The higher scaling factor in late spiral and irregular galaxies can be attributed to their lower metallicity with respect to early spiral galaxies \citep[e.g.,][]{Gonzalez15}: as discussed in \S\ref{txt:discmetal} low metallicity galaxies show an excess of ULXs. \autoref{fig:metalvst} shows the metallicity distribution for different morphological types in the \hec{} and our host galaxy sample. We see that the average metallicity quickly drops for galaxies later than Sc, the same galaxies for which the scaling factor (see \autoref{tbl:earlylate}) increases from $0.51$ to $1.54\aunits{}$. However, these trends do not account for another important factor which cannot be tested in the current sample: late-type galaxies are more prone to short and intense star-formation episodes, which might increase their ULX content significantly \citep[e.g.,][]{Wiktorowicz17}, although the effect of metallicity appears to have a stronger effect in the X-ray output of a galaxy \citep[e.g.,][]{Fragos13a}.

In order to account for the LMXB contribution in LTGs, in \S\ref{txt:earlylate} we computed the scaling parameters $\alpha$ and $\beta$ for the linear relation between number of observed sources with $L_{\rm X}{>}10^{39}\ergs$ and both SFR and \mass{} (\autoref{eq:modelab}). The value of $\alpha{=}0.45^{+0.06}_{-0.09}\aunits{}$ (see \autoref{tbl:2dfits}) for all LTGs is somewhat smaller, but consistent with the value of $a{=}0.51{\pm}0.06\aunits{}$ found using the model of \autoref{eq:modela} where only the SFR scaling is considered (see \autoref{tbl:earlylate}). The smaller scaling when accounting for the contribution of the \mass{} is the result of the small fraction of the ULX population that is associated with the old stellar population (and consequently the \mass{}) in spiral galaxies. The results of the fits for early and late spirals (see \autoref{tbl:2dfits} and \autoref{fig:2dfits_nAGN}) illustrate that the \mass{} contribution is significant in early spirals (S0/a-Sbc) at the $2\sigma$-level, while it can be neglected in late spirals ($\beta{<}6.6\bunits{}$ with most probable value 0.0).

Recently, \citet{Lehmer19} constructed luminosity functions of XRBs as a function of both SFR and \mass{} to account for the contribution of both LMXBs and HMXBs. Using the best-fitting parameters for their full sample (see table~4 in \citealt{Lehmer19}) we integrated the LF above the ULX limit (\ulxlimit). We find that they predict
$N_{\rm ulx}=a_{\rm L19} {\rm SFR}+b_{\rm L19} \mass{}$, where $\alpha_{\rm L19} = 0.62{\pm}0.08\aunits{}$ and $\beta_{\rm L19}=18^{+23}_{-11}\bunits{}$.
The scaling with SFR ($\alpha$) is consistent at the $1\sigma$-level with our findings for all LTG galaxies ($0.45^{+0.06}_{-0.09}\aunits{}$; see \autoref{eq:modelab}). The scaling with \mass{} ($\beta$) is highly uncertain in the ULX regime, but also consistent at the $1\sigma$-level with the one we find for all LTG galaxies ($3.3^{+3.8}_{-3.2}\bunits{}$).

Finally, the above results are consistent with the qualitative picture shown in \autoref{fig:msfrplane}; the number of ULXs in LTGs increases with both SFR and \mass{}. Note that the trend of galaxies hosting larger population of ULXs at higher sSFR (see diagonal lines), may have a trivial explanation: ULXs being primarily associated with young stellar populations, are more abundant in galaxies with higher SFR and/or lower mass. However, an age effect may be at the play: starbursts have high sSFR, by definition, and are expected to have high formation rate of BH ULXs, which dominate the population at ${\sim}5\,\rm Myr$ \citep[e.g.][]{Fragos13b}.

\begin{figure}
    \centering
    \includegraphics[width=\columnwidth]{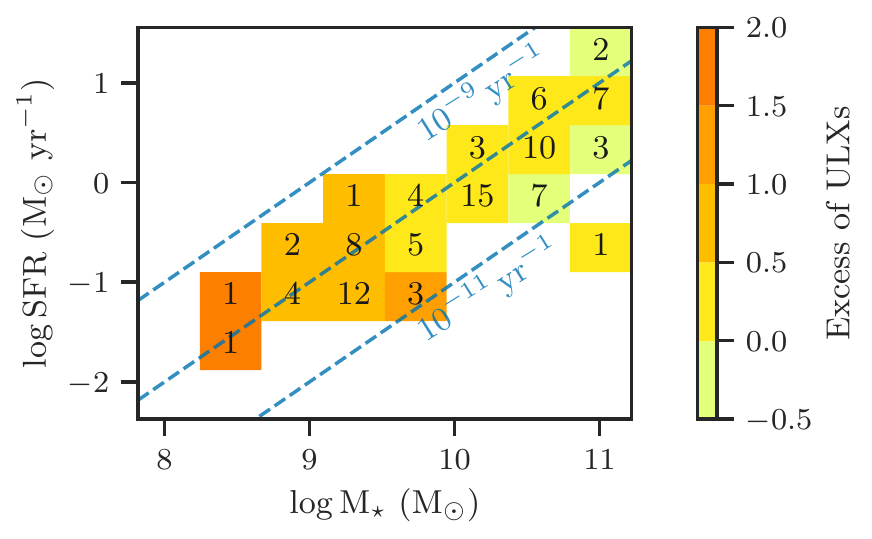}
    \caption{The mean excess of ULXs (colour scale; as expressed in \autoref{eq:excess}) in LTGs as a function of SFR and \mass{}. The diagonal lines correspond to indicative specific SFRs. The numbers in the boxes denote the number of galaxies in each bin. Despite their small numbers, the data show that galaxies with low masses exhibit a clear excess in the number of ULXs over the expectation.}
    \label{fig:msfrplanenormed}
\end{figure}

Such an age effect will manifest as an excess of ULXs in high sSFR galaxies compared to the expectation from the average SFR-\mass{} scaling relation based on all LTGs in our sample.
We assess this by defining the \emph{excess} of ULXs,
\begin{equation}
    \text{excess} = \log \frac{N_{\rm obs}}{N_{\rm exp}}
    \label{eq:excess}
\end{equation}
where $N_{\rm obs}$ is the number of ULX candidates, and $N_{\rm exp}$ is the expected number of sources according to the model in \autoref{eq:modelab} and its best-fitting values (\autoref{tbl:2dfits}). In order to explore the possible dependence of the ULX excess on SFR and \mass{}, we plot in \autoref{fig:msfrplanenormed} the ULX excess of galaxies as a function of their SFR and \mass{}. We do not see a dependence on the sSFR; instead, it is clear that low-mass galaxies present an excess of ULXs, in agreement with \citet{Swartz08}. Since low-mass galaxies tend to present lower metallicities \citep[e.g.,][]{Kewley08} we interpret this excess as likely being related to the metallicity of their hosts.

\subsection{Excess of ULXs in low-metallicity galaxies}
\label{txt:discmetal}

There is a growing observational body of evidence for an excess of ULXs in low-metallicity galaxies \citep[e.g.,][]{Soria05,Mapelli10,Prestwich13,Brorby14,Tzanavaris16}. This trend can be interpreted theoretically in the context of the weaker stellar winds in low-metallicity stars. The stars retain higher fraction of their initial mass, and as a consequence, more massive BHs are formed, with smaller orbital separation due to weaker angular momentum losses (e.g., \citealt{Heger03,Belczynski10,Marchant17}). In addition, the tighter orbits result in an increased fraction of HMXBs that enter a Roche-lobe overflow phase, which being a more efficient accretion mechanism than stellar winds, leads to more luminous X-ray sources \citep{Linden10}.

\begin{figure*}
    \centering
    \includegraphics[width=\textwidth]{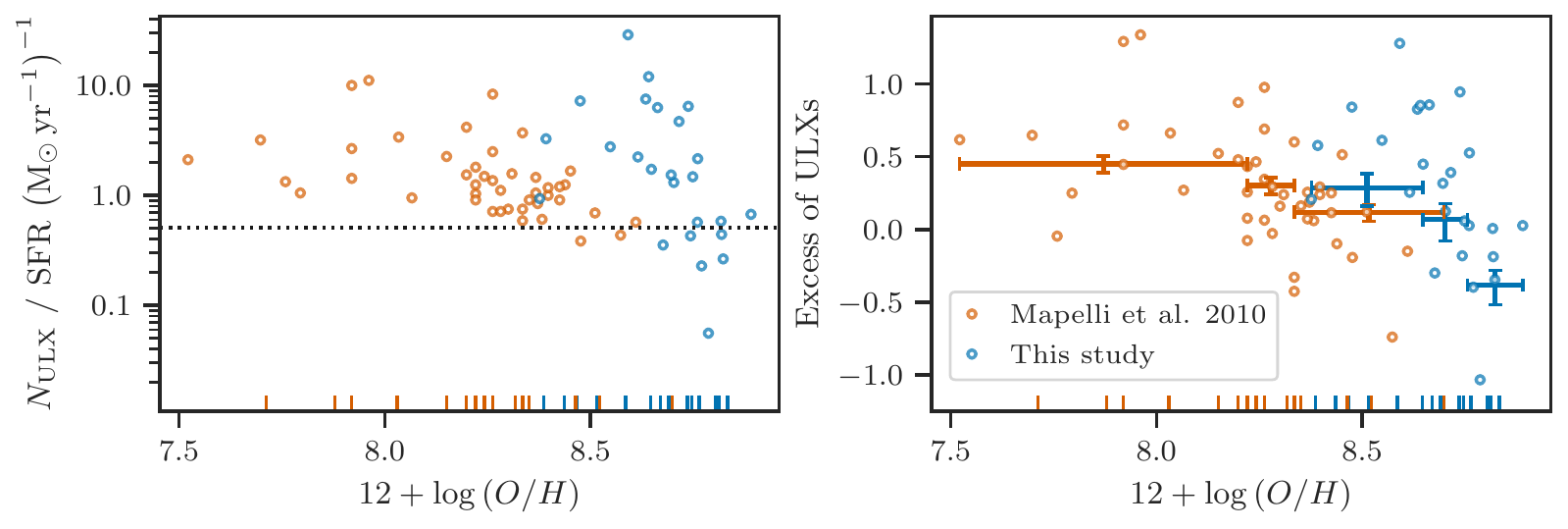}
    \caption{Left panel: the number of ULXs per SFR unit ULXs for LTGs in our sample (blue) with metallicity estimates (blue points), and that of \citet{Mapelli10} (orange). The metallicities of the galaxies with no sources above the ULX limit, are shown as ticks on the $x$-axis. Low-metallicity galaxies present an excess of ULXs with respect to the average scaling (dotted horizontal line.)
    Right panel: same as the left panel but now the $y$-axis is the excess of ULXs as defined in \autoref{eq:excess}.  We find a significant anti-correlation (Kendall rank correlation coefficient $\tau{=}{-}0.43$ with $p$-value $0.002$). We bin the galaxies to reduce the stochasticity of the ULX excess (see \S\ref{txt:discmetal}) and compute the median and 68\% CIs of the excess as a function of the metallicity (blue error bars). We repeat the same procedure for the ULX excess in the sample of \citet{Mapelli10} (orange) which also exhibits significant anti-correlation ( $\tau{=}{-}0.40$ and $p{=}0.001$).
    \label{fig:excess}}
\end{figure*}

To investigate the correlation of the ULX population with metallicity, we plot in \autoref{fig:excess} (left) the number of ULXs per SFR as function of the host galaxy metallicity for the 44 galaxies with metallicity measurements in the \hec{} (blue circles). We see an excess of ULXs in low metallicities with respect to the average relation shown with the dotted line. In the right panel of  \autoref{fig:excess} we plot the  ULX excess (\autoref{eq:excess}) against metallicity. We see that the frequency of ULXs indeed increases with decreasing metallicity (Kendall rank correlation coefficient $\tau{=}-0.43$ with $p$-value of $0.002$). For comparison, we also plot in the same figure the excess of ULXs computed from the sample in \citet{Mapelli10} (orange circles) using their reported values for (i) the number of ULXs, (ii) the expected background contamination, and (iii) the SFRs of the host galaxies\footnote{Since \citet{Mapelli10} do not provide \mass{} estimates, which are needed to compute $N_{\rm exp}$ in \autoref{eq:excess}, we obtain our own estimates for $N_{\rm exp}$ using the SFR and the scaling constant of $0.51\aunits{}$, determined from fits presented in \S\ref{txt:earlylate} (cf. \autoref{eq:modela}). In addition, metallicities from \citet{Mapelli10} were converted from solar units ($\rm Z_\odot$) using their adopted solar metallicity  $12+\log\left(O/H\right)_\odot{=}8.92$.}. To reduce the Poisson noise, the galaxies are grouped in metallicity bins (defined to have a similar number of objects in each bin and always more than eight) shown as $x$-axis error bars in \autoref{fig:excess}. The central values and the error bar length in the $y$-axis correspond to the median and the 68\% confidence interval of the ULX excess, computed by accounting for Poisson uncertainty of the number of sources and interlopers.

Based on the binned statistics in \autoref{fig:excess}, we find that the galaxies with the lowest metallicities in our sample (corresponding to $0.3$-$0.5\,\rm Z_\odot$) host more ULXs per SFR by a factor of ${\sim}2$, in comparison to galaxies of intermediate metallicity ($0.5$-$0.7\,\rm Z_\odot$) which present no excess of ULXs. Interestingly, galaxies with near-solar metallicity (${>}0.7\rm\,Z_\odot$) present a deficiency of ULXs; they host half of the expected ULX population.

The same trend is observed in the sample of \citet{Mapelli10}. However, there seems to be a small horizontal offset of ${\sim}0.25 \rm\, dex$ between our study and that of \citet{Mapelli10}. We attribute this offset to the different metallicity calibrations\footnote{In our sample we use the metallicity estimates in the \hec{} which were calculated via the \ion{O}{III}-\ion{N}{II} calibration in \citet{PP04}, while the metallicities in \citet{Mapelli10} are based on many different calibrations, mainly those in \citealt{Pilyugin01} and \citealt{Pilyugin05}.} which can have systematic biases up to $0.7\rm\,dex$ (see fig.~2 in \citealt{Kewley08}). Using eight common galaxies in our sample and that of \citet{Mapelli10}, we find that the mean offset between the metallicities is $0.28\pm0.09$.

In conclusion, an excess of ULXs is linked with low-metallicities. This is also in line with our result that the ULX-SFR scaling factor is significantly higher in later-type galaxies, for which the metallicity is lower (see \S\ref{txt:dislate}). This excess has direct implications for the XRB content of the high-redshift Universe. The mean metallicity of galaxies at $z{\sim}2.5$ was only ${\sim}0.1\,\rm Z_\odot$ \citep[e.g.,][]{Madau14}.
Indeed, an excess of the integrated X-ray luminosity per unit SFR is seen in observational studies of high-redshift galaxies \citep[e.g.,][]{Lehmer05,Lehmer16,Basu13a,Basu13b,Basu16,Brorby16,Fornasini19,Fornasini20,Svoboda19}.  Our results indicate that this excess is the result of a larger population of luminous X-ray sources per unit SFR in lower metallicities.
However, we cannot exclude the possibility that the stellar population age also plays a role on the ULX excess. Since the metallicity and age can vary by region in a galaxy, investigation on sub-galactic scales can help to disentangle their relative effects on the XRB populations (cf. \citealt{Anastasopoulou19,Lehmer19,Kouroumpatzakis20}).

\subsection{ULXs and old stellar populations}
\label{txt:disearly}

ULXs in elliptical galaxies \citep[e.g.,][]{Laurence05} are considered to belong to the high-end of the LMXB-LF \citep[e.g.,][]{Swartz04,Plotkin14}. Notably, rejuvenation of stellar populations due to galaxy mergers might also produce additional ULXs \citep{Zezas03,Raychaudhury08,Kim10}. In addition, it is possible that a small population of ULXs are dynamically formed in GCs \citep[e.g.,][]{Maccarone07,Dage20}. Indeed, we find evidence that a small but significant population of ULXs in elliptical galaxies resides at large galactocentric distances (see \S\ref{txt:spatial}), i.e., not following the \mass{} distributions. Literature review of the hosts of these sources showed evidence for recent merger activity, or large GC populations, indicating that these ULXs could be associated with GCs, given the flatter distributions of GCs and their LMXB populations with respect to the stellar light \citep[e.g.,][]{Kim06}.

In \S\ref{txt:earlylate}, based on the fit of the number of ULXs against the \mass{} of ETGs (see \autoref{eq:modelb}), we  find $15.1^{+3.9}_{-3.6}\bunits{}$ in the non-AGN sample. However, it is higher by a factor of $2$-$3$ than the expectation from the LMXB-LF of \citet{Zhang12} ($5\pm2 \bunits$), and the specific ULX frequency in \citet{Plotkin14} ($6.2\pm1.3 \bunits$) and \citet{Walton11} (${\sim}7\bunits$). While these studies address possible contamination from AGN in their X-ray source samples, they still consider (except for \citealt{Zhang12}) the total $K$-band luminosity of the galaxies as a tracer of the \mass{} even if the galaxy hosts an AGN. The contamination by the AGN would lead to an overestimation of the \mass{}, and consequently an underestimation of the specific ULX frequency. To quantify this effect, in Appendix~\ref{app:withagn} we compute the specific ULX frequency in the full sample (including galaxies hosting AGN, but still excluding their nuclear sources). We find $6.3^{+1.0}_{-0.9}\bunits{}$ in good agreement with the literature estimates.

Why do non-AGN ETGs exhibit higher specific ULX frequency than the \full{} ETG sample? As we show in Appendix~\ref{app:withagn} the presence of an AGN does not significantly affect the observed $K$-band luminosity and therefore the measured \mass{}. The AGN contribution is ${<}10\%$ of the total $K$-band luminosity (Bonfini et al. 2020, submitted). However, the full sample extends to much larger masses than the non-AGN ETG sample. Consequently, the difference in the specific ULX frequency between the full and non-AGN sample could be explained by a non-linear dependence of the number of ULXs on the \mass{}.

\begin{figure}
    \centering
    \includegraphics[width=\columnwidth]{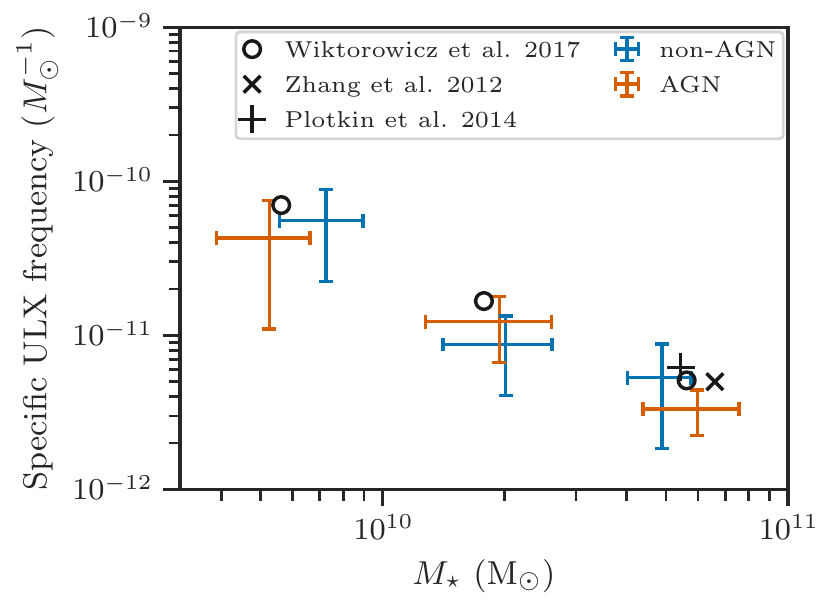}
    \caption{The specific ULX frequency for ETGs in our sample with stellar masses in the ranges $10^{9.5}$-$10^{10}$, $10^{10}$-$10^{10.5}$ and $10^{10.5}$-$10^{11}\,\rm M_\odot$, for AGN (orange error bars) and non-AGN galaxies (blue error bars). The cross and plus markers indicate the specific ULX frequency and the mean stellar mass of the samples in \citet{Zhang12} and \citet{Plotkin14}. The circles indicate the computed values by convolving the binary population synthesis results of \citet{Wiktorowicz17} with average SFH of ETGs in \citet{McDermid15}. Note that the $x$-axis error-bars do not indicate bin widths, but the standard deviation of the stellar masses of the galaxies contributing in each bin, to give a sense of the stellar mass distribution in each bin.}
    \label{fig:specific}
\end{figure}

In order to quantify the dependence of the specific ULX frequency on the \mass{}, we compute the scaling factor $b$ in our ETG sample, over three \mass{} bins ($10^{9.5}-10^{10.0}$; $10^{10.0}-10^{10.5}$; $10^{10.5}-10^{11.0}\,\rm M_\odot$), separately for AGN and non-AGN galaxies. This is shown in \autoref{fig:specific}. The results for the AGN and non-AGN samples agree within the errors, as expected based on the previous assessment that the AGN do not lead to significant overestimation of the \mass{} (see Appendix~\ref{app:withagn}). They also agree with the scalings reported in \citet{Zhang12} and \citet{Plotkin14} for the corresponding \mass{} bins (also plotted in \autoref{fig:specific}). Interestingly, however, we find that $b$ depends strongly on the \mass{} of the host galaxy. This dependence explains the lower specific ULX frequency found in the \full{} ETG sample which is biased towards more massive galaxies than in the case of non-AGN ETG sample.

The dependence of the specific ULX frequency on the \mass{} could be caused by star-formation history (SFH) differences in ETGs \citep{McDermid15}. Simulations indicate that ULXs with neutron-star accretors and red giant or Hertzsprung-gap donors can appear several hundreds of Myrs after a star-formation episode \citep{Wiktorowicz17}, and therefore their frequency in early-type galaxies is expected to be strongly dependent on the SFHs. Calculated specific ULX frequencies (circles in \autoref{fig:specific}) using the \citet{Wiktorowicz17} simulation and the \citet{McDermid15} average SFHs for the same stellar mass ranges (see \S\ref{txt:models}), are in excellent agreement with the observed specific ULX frequencies in our sample.

Therefore, comparisons of ULX rates in ETGs should account for the mass range covered in each sample and the corresponding bias due to different SFHs. In this respect, we attribute differences between our estimates of the specific ULX frequency and those of previous studies to the different \mass{} ranges in the samples. Note, however, that the specific ULX frequency was found to be constant in elliptical galaxies in \citet{Walton11}, albeit with a relatively small sample of 22 galaxies.

Furthermore, in \S\ref{txt:rateulx} we find that lenticular galaxies in our sample host 2-3 times fewer ULXs than elliptical galaxies, by a factor of $2$-$3$, even when normalising by the \mass{}. This result is in agreement with the findings of \citet{Wang16}. However, we noticed that this difference disappears when considering the \full{} sample. Given the dependence of the specific ULX frequency on the \mass{} shown in \autoref{fig:specific}, it is possible that this discrepancy stems from the different \mass{} regimes of the corresponding samples. Indeed, we find that the interquartile range (middle 50\%) of the stellar masses in the \full{} sample lies in the $1.6{\times}10^{10}\massunit{}$-$10^{11}\massunit{}$ range, for both elliptical and lenticular galaxies, while in the case of the non-AGN sample, lenticular galaxies present higher masses, $1.4$-$4{\times}10^{10}\massunit{}$, compared to that of the elliptical galaxies, $0.5$-$3.5{\times}10^{10}\massunit{}$.

\subsection{Comparison with models}
\label{txt:models}

Comparison of binary population synthesis models and demographic studies of ULXs provide tests for models of the formation and evolution of X-ray binaries with extreme mass transfer rates.

We compare our findings with the results in \citet{Wiktorowicz17}, who computed the observed number of ULXs as a function of time for three different metallicities (0.01, 0.1 and 1 $Z_\odot$). Since the SFR indicators used in our study are based on the IR emission which is sensitive to stellar populations of ages up to ${\sim}100\rm\,Myr$ \citep{Kennicutt12}, we compare our results with the number of ULXs reported in \citet{Wiktorowicz17} observed after $100\,\rm Myr$ for a starburst scenario for $6\times10^{10}\,\rm M_\odot$ of stars formed with $100\,\rm Myr$ duration. They report $4\times10^{2}$ ULXs which corresponds to a formation rate of $0.67\aunits$. This value is close to our results for all LTGs ($0.51{\pm}0.06\aunits{}$; see \autoref{tbl:earlylate}).

To study the effect of metallicity, we also consider the $0.1\,Z_\odot$ simulation from \citet{Wiktorowicz17}. The resulting formation rate is $12\aunits{}$, about 18 times stronger than that in the case of $Z{=}Z_\odot$. As shown in \autoref{fig:excess}, our sample at such low metallicities is insufficient to estimate the excess of ULXs. However, we find an excess of ${\sim}0.7\rm\,dex$ in the ULX rate for ${\sim}0.3\rm\,dex$ lower metallicities in comparison to the bulk of the galaxies (which are predominantly solar metallicity galaxies). This translates to a factor of ${\sim}5$ more ULXs at $Z{=}0.3 Z_\odot$ which is between the expectations from the models of \citet{Wiktorowicz17} for $Z{=}0.1\,Z_\odot$ and $Z{=}Z_\odot$.

\begin{figure}
    \centering
    \includegraphics[width=\columnwidth]{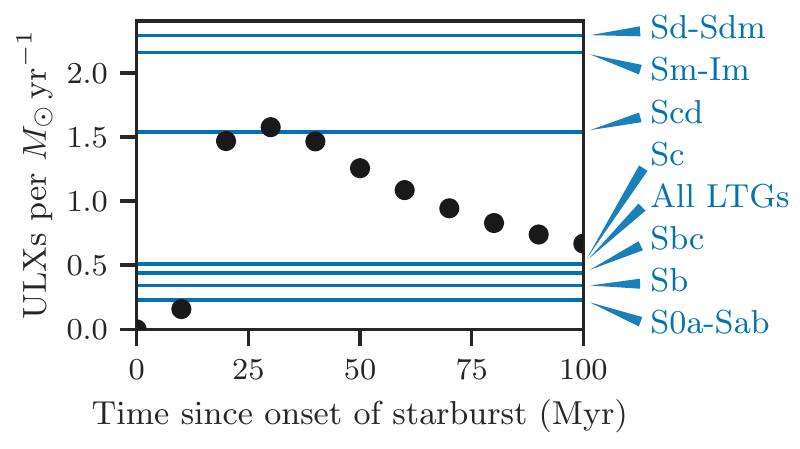}
    \caption{Formation rate of ULXs as a function of time since the onset of a star-formation episode (with constant SFR) based on the simulations of \citet{Wiktorowicz17} for solar metallicity. The horizontal lines indicate the parameter $a$ (number of ULXs per SFR) found in different morphological types of LTGs in our sample (see \autoref{tbl:earlylate}). The range of $a$ in LTGs is comparable to the range of the computed formation rate of ULXs at timescales of $10-100\rm\,Myr$, except for Sd-Sdm and Sm-Im galaxies, possibly because of their sub-solar metallicity (see \S\S\ref{txt:dislate}, \ref{txt:discmetal}).}
    \label{fig:agewik}
\end{figure}

However, the SFHs of real galaxies may present individual star-formation episodes (as can be the case in irregular galaxies). This is expected to have a strong effect on the formation rate of XRBs, as it has been demonstrated in HMXB populations \citep[e.g.,][]{Antoniou16,Antoniou19,Lehmer19}, and the observed populations of ULXs which are typically associated to recent star-formation episodes. Although the SFHs of the galaxies in our sample are not known, based on the simulations of \citet{Wiktorowicz17} we can estimate the range of ULX formation rates as a function of time in a continuous SF episode over a time-scale of 100\,Myr. We compute the formation rate as the number of ULXs at time $t$ from fig.~2 in \citet{Wiktorowicz17} divided by the stellar mass of the parent population formed in the same SF episode. By performing this computation for various values of $t{\leq}100\rm\,Myr$, we find formation rates in the range $0 - 1.58\aunits{}$ (see \autoref{fig:agewik}) close to the range of the ULX-SFR scaling in samples of different morphological classes (see \autoref{tbl:earlylate}), except for the late-type spiral galaxies and irregulars (Sd-Im). The latter present an excess of ULXs due to their lower metallicities (see \S\ref{txt:dislate}).

In the case of ETGs, most of the ULXs are expected to be long-lived systems of LMXBs with ages ${\gg}100\,\rm Myr$, but as shown in \citet{Wiktorowicz17}, the number of ULXs decreases with time since the SF episode. Therefore, the number of observed ULXs in ETGs, depends strongly on the SFH. In addition, recently, it has been shown in \citet{McDermid15} that the age of stellar populations in ETGs can vary more than it was thought before. In order to compute a fiducial range of specific ULX frequency in ETGs, we use the average SFH of ETGs in \citet{McDermid15} in three stellar-mass ranges: $\log{{\rm M_\star}}\in \left(9.5, 10.0\right)$, $\left(10.0, 10.5\right)$ and $\left(10.5, 11.0\right)$. These ranges cover the majority of the ETGs in our sample (${\sim}90\%$).

For the three average SFHs, we compute the number of expected ULXs at the present time by convolving the SFHs (cf. \citealt{Kouroumpatzakis20}) with the ULXs rates\footnote{Since \citet{Wiktorowicz17} do not provide an instantaneous SB response function, rather a SB of duration of 100\,Myr, the convolution is performed in bins of 100\,Myr.} per unit stellar mass from the prediction of \citet{Wiktorowicz17} for solar metallicity. Then, we divide by the the midpoint for each mass range in order to calculate the specific ULX frequencies in each mass range. We find 70.1, 16.7 and 5.1$\bunits{}$ for the low-, intermediate- and high-mass ETGs, suggesting that the ULX content of ETGs is indeed a strong function of SFH (see \autoref{fig:specific}). These estimates are comparable to the rates we derived from our ETGs sample, $b{=}15.1^{+3.9}_{-3.6}\bunits{}$ for non-AGN ETGs, and $b{=}6.3^{+1.0}_{-0.9}\bunits{}$ for the full ETGs sample (which is biased towards the higher mass bin; see \autoref{tbl:earlylate} and Appendix~\ref{app:withagn}).

\subsection{Limitations of this study}

The parent sample of the \hec{}, the \hyper{}, includes galaxies and measurements from a multitude of surveys with different sky coverage and sensitivity. Similarly, parameters provided in \hec{} (e.g., SFR, \mass{}, metallicity, AGN classifications) are derived from combinations of data from all-sky surveys (e.g. \iras{}, \twomass{}) and the \sdss{} (e.g., M/L ratios, \wise{} forced photometry of \sdss{} objects). Despite the unknown selection function of the parent sample, the \hec{} is the most complete sample of galaxies in the local Universe with available information on their stellar content, allowing us to draw meaningful conclusions regarding the ULX scaling relations covering a very broad stellar mass ($10^7.5$-$10^{11.5}\,\rm M_\odot$) and SFR range ($10^{-2.5}$-$10^{2}\sfrunit{}$; \S\ref{txt:representativeness}).

Similarly, the serendipitous nature of the \CSC{} leads by definition to a non-uniform X-ray sample. In addition, to avoid contamination from X-ray emitting AGN, we exclude nuclear sources in galaxies that were either classified as AGN or we did not have information on their nuclear activity. The drawback of this approach is that we may have excluded circum-nuclear ULXs.

Finally, for the study of scaling relations, we primarily consider a secure sample of non-active galaxies to avoid the overestimation of SFR and \mass{} due to nuclear activity. This practice reduces the sample used for the ULX investigations, and may have removed known bona-fide starforming ULX hosts (e.g., Holmberg II). In addition, it leads to a bias against massive galaxies which are more likely to be targeted as AGN hosts. As discussed in \S\ref{txt:disearly}, including the AGN sample, at least in the ETGs, does not bias the measured galaxy properties.

\section{Summary}
\label{txt:summary}

We construct a census of ULXs in nearby galaxies ($D{<}40\,\Mpc{}$) by cross-matching the \CSC{} and the \hec{}. We use this sample in order to study the ULX rates as a function of morphology, SFR, \mass{} and metallicity of their host galaxies. We deliver a sample of host galaxies and their ULX populations that serves as a benchmark for models describing the nature, formation and evolution of ULXs. We
\begin{enumerate}
    \item constrain the number of ULXs in LTGs as a function of SFR, and both SFR and \mass{} (to account for the LMXB contribution):
    \begin{align}
        N_{\rm ulx} &= \left(0.51 \pm 0.06\right) \times \dfrac{\rm SFR}{\sfrunit{}} \\
        N_{\rm ulx} &= 0.45^{+0.06}_{-0.09} \times \dfrac{\rm SFR}{\sfrunit} + 3.3^{+3.8}_{-3.2} \times \dfrac{\mass{}}{10^{12}\,\rm M_\odot}.
    \end{align}
    \item find that the ULX-SFR scaling increases with the morphological type of LTGs.
    \item verify the excess of ULXs in low-metallicity galaxies, which partially drives the above mentioned trends with the morphological type.
    \item find evidence for evolution of the specific ULX frequency in ETGs with their \mass{}, which we attribute to their different SFHs.
    \item find that our observed scaling relations can be reproduced by published ULX formation rates from population synthesis models when accounting for the galaxies SFHs and/or metallicity.
\end{enumerate}

While {\it eROSITA} \citep{Predehl10,Merloni12} will provide a uniform flux-limited sample of normal galaxies and ULXs in the local Universe \citep[e.g.,][]{Basu20}, serendipitous surveys with \chandra{} will continue to probe unconfused ULX populations at larger distances and their connection to the lower luminosity XRB populations.

\section*{Acknowledgements}

The authors would like to thank the anonymous referee for providing helpful comments that improved the paper.
KK thanks R. D'Abrusco, K. Anastasopoulou, P. Bonfini, F. Civano, G. Fabbiano, K. Kouroumpatzakis, A. Rots and P. Sell for their suggestions.
The scientific results reported in this article are based to a significant degree on data obtained from the {\it Chandra Data Archive} and the {\it Chandra Source Catalog} provided by the {\it Chandra X-ray Center (CXC)}.
The research leading to these results has received funding from the
{\it European Research Council} under the European Union's
{\it Seventh Framework Programme} (FP/2007-2013) /
{\it ERC} Grant Agreement n.~617001,
and the {\it European Union’s Horizon 2020} research and innovation programme under the {\it Marie Sk\l{}odowska-Curie RISE} action, Grant Agreement n.~691164 ({\it ASTROSTAT}).
BDL acknowledges support from the {\it NASA Astrophysics Data Analysis Program} 80NSSC20K0444.

\section*{Data availability}

The data underlying this article are available in the article and in its online supplementary material.


\bibliographystyle{mnras}
\bibliography{main}


\appendix

\section{Galactocentric scale parameter}
\label{app:ellipses}

We define the galactocentric scale parameter, $c$, as the deprojected distance of a source from the centre of its host galaxy, normalised by the galaxy's semi-major axis. Since the shapes of the galaxies in our study are defined through isophotal ellipses and the length of the semi-major axis is free of projection effects, we observe that $c$ can be computed as the ratio of the semi-major axes of two projected ellipses: a scaled version of the isophotal ellipse (same centre, orientation and axis ratio) passing through the source, and the isophotal ellipse itself.
Consequently, for a source at $\left(\alpha_s, \delta_s\right)$ and a galaxy centred at $\left(\alpha_g, \delta_g\right)$ with semi-axes $R_1, R_2$ and position angle $\omega$ measured from North to East, the scale $c$ is found by
\begin{enumerate}
    \item
        rotating the coordinates so that the centre of the galaxy falls in $(\alpha, \delta){=}\left(0, 0\right)$ and the semi-major axis is on a meridian ($a{=}0$)
    \item
        setting the sum of the great-circle distances of the source from the focal points to be equal to two times the semi-major axis (of the scaled version of the ellipse)
\end{enumerate}

Step (i) is performed by converting the spherical coordinates to Cartesian (unit radius):
\begin{equation*}
    \left(x, y, z\right) = \left(\cos\delta_s \cos\alpha_s, \cos\delta_s \sin\alpha_s, \sin\delta_s \right),
\end{equation*}
and rotating around the $z$-axis by $-\alpha_g$, the $y$-axis by $\delta_g$ and the $x$-axis by $-\omega$ (to align the semi-major axis with the meridian), by multiplying with the corresponding 3D-rotation matrices:
\begin{equation*}
    \begin{pmatrix} x' \\ y' \\ z' \end{pmatrix} = \mathbf{R}_x(-\omega) \cdot \mathbf{R}_y(\delta_g) \cdot \mathbf{R}_z(-\alpha_g) \cdot
    \begin{pmatrix} x \\ y \\ z \end{pmatrix}.
\end{equation*}
The final coordinates are converted back to spherical coordinates:
\begin{equation*}
    \left(\alpha, \delta\right) = \left(
        \tan^{-1}\frac{y'}{x'},
        \frac{\pi}{2} - \cos^{-1}z'
    \right).
\end{equation*}

\begin{figure}
    \centering
    \begin{minipage}[b]{0.45\columnwidth}
        \includegraphics[width=\textwidth]{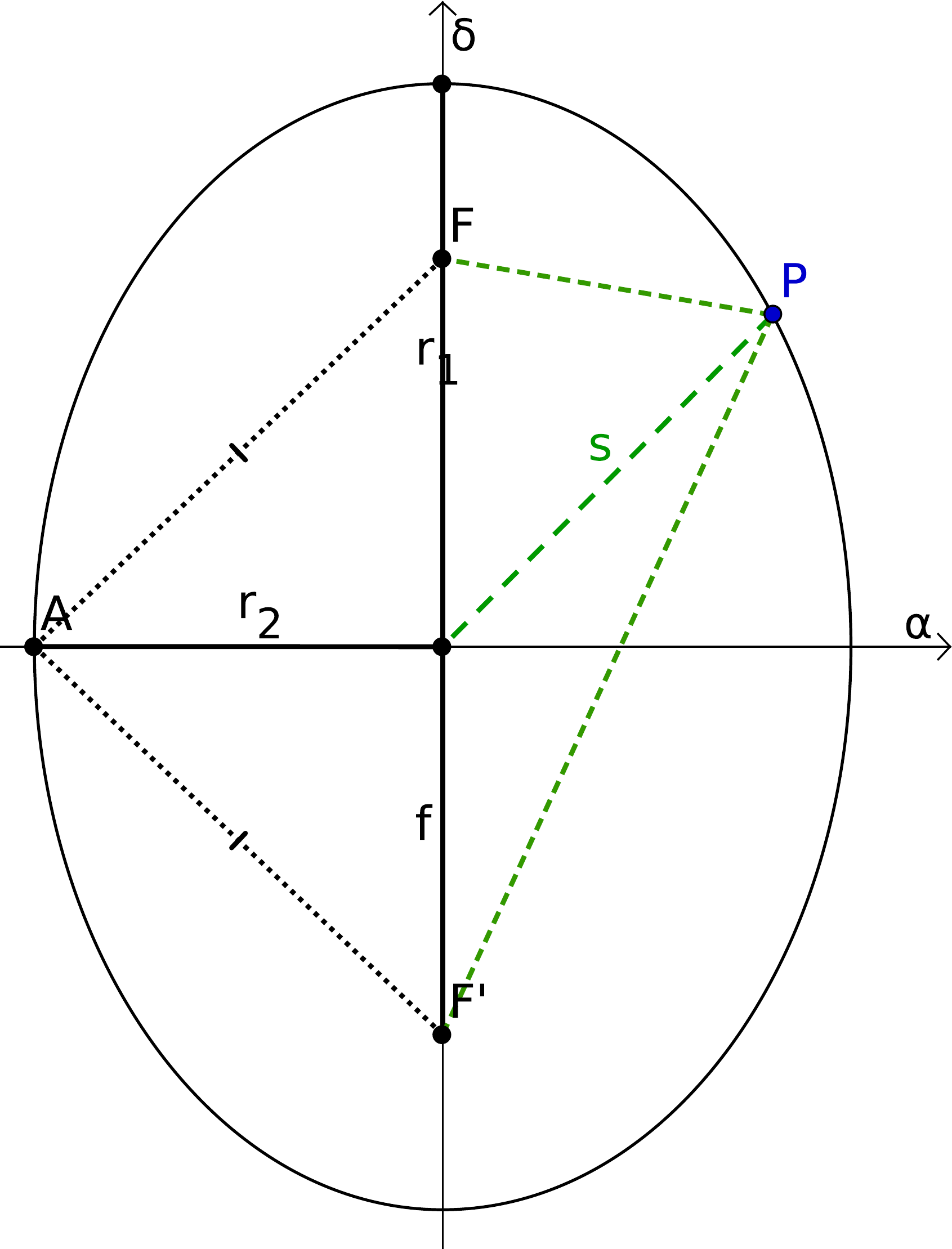}
    \end{minipage}
    \hfill
    \begin{minipage}[b]{0.47\columnwidth}
        \caption{The scaled version of the ellipse passing through the source ($P$). The separation of the source $s$ and its distance from the two focal points ($F$, $F'$) is denoted with green dashed lines. The solid black lines denote the semi-major and semi-minor axes, while the dashed black lines indicate the distance of the co-vertex ($A$) to the focal points. The co-vertex $A$ is introduced so that the focal distance $f$ is estimated in an intermediate step.}
        \label{fig:ellipse}
    \end{minipage}
\end{figure}

Step (ii) consists of finding the parameters of an ellipse shown in \autoref{fig:ellipse} for which the semi-major and semi-minor axes are scaled versions of the original ellipse $\left(r_1, r_2\right){=}\left(c R_1, c R_2\right)$. This is done by requiring $\left(F,P\right) + \left(F',P\right){=}\left(F,A\right) + \left(F',A\right)$ where $(A, B)$ denotes the great-circle distance between points $A\left(\alpha_1, \delta_1\right)$ and $B\left(\alpha_2, \delta_2\right)$, computed by employing a form of the Haversine formula which is more precise for nearby points:
\begin{equation*}
    s = 2 \sin^{-1}\sqrt{\sin^2\frac{\delta_1 - \delta_2}{2} + \cos\delta_1 \cos\delta_2 \sin^2\frac{\alpha_1 - \alpha_2}{2} }.
\end{equation*}

The solution in terms of $c$ (the scale of the ellipse) is found by solving for $c$ the equation:
\begin{equation}
    \sin^{-1}\sqrt{u^{-} + w} + \sin^{-1}\sqrt{u^{+}+w} = c R_1,
    \label{eq:gcnumerical}
\end{equation}
where
\begin{align*}
    u^{\pm} = \sin^2\frac{f{\pm}\delta}{2},
    w = \cos f \cos\delta \sin^2\frac{\alpha}{2},
    f = \cos^{-1}\left[\frac{\cos\left(c R_1\right)}{\cos\left(c R_2\right)}\right].
\end{align*}
Since the \autoref{eq:gcnumerical} is not in closed form, it is solved numerically. Due to the periodicity of trigonometric functions, there are multiple solutions corresponding to ellipses engulfing the celestial sphere multiple times. To avoid this, we require that $r_2$ is less than $\frac{\pi}{2}$. Also, the separation of the source from the centre of the galaxy acts as a lower and upper limit for the semi-major and semi-minor axes respectively. Therefore the galactocentric distance is constrained on
\begin{equation*}
    c \in \left[\frac{s}{R_1}, \min\left\{\frac{s}{R_2},
                                          \frac{\pi}{2 R_1}\right\} \right],
    \label{eq:gcrange}
\end{equation*}
where $s$ is now expressed in the transformed coordinates as:
\begin{equation*}
    s = 2 \sin^{-1}\sqrt{\sin^2\frac{\delta}{2} + \cos{\delta}\sin^2{\frac{\alpha}{2}}}.
\end{equation*}

\section{Results of fits with and without AGN hosts}
\label{app:withagn}

\begin{figure}
    \includegraphics[width=\columnwidth]{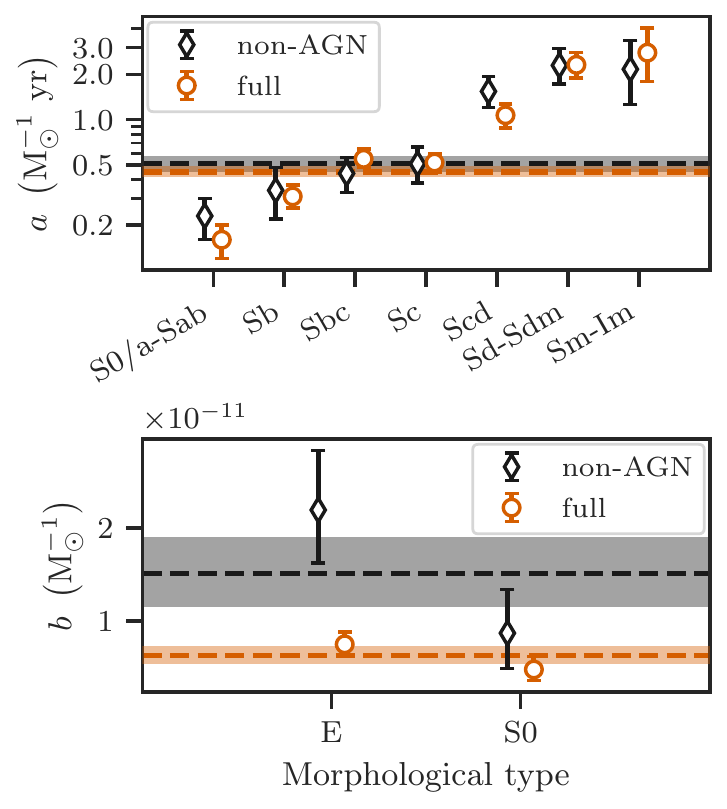}
    \caption{Comparison of fitting results between non-AGN and \full{} samples.
    \textbf{Top}: the scaling parameter $a$ (see \autoref{eq:modela}) in non-AGN LTGs (dashed black line) and its 68\% CI (grey band), and in the \full{} sample (orange line and band). Fitting results for various morphological subclasses of LTGs are shown with errorbars.
    \textbf{Bottom}: same as top panel, but now for the scaling parameter $b$ (see \autoref{eq:modelb}) in ETGs and separately in elliptical and lenticular galaxies.}
    \label{fig:earlylatefits}
\end{figure}

\begin{figure}
    \includegraphics[width=\columnwidth]{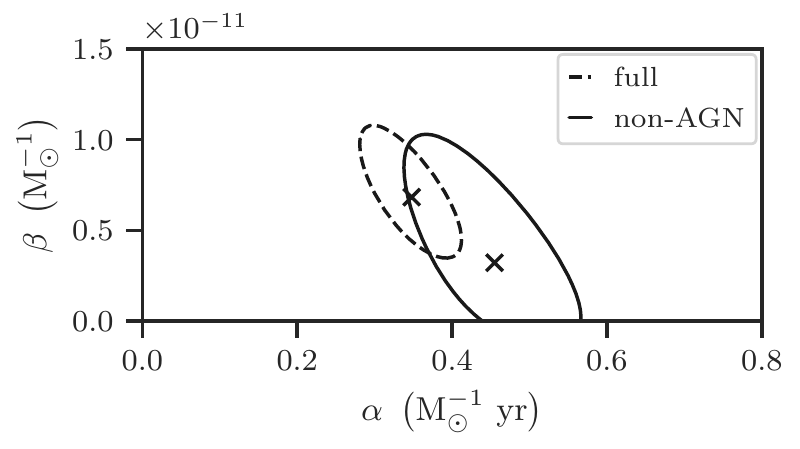}
    \caption{The best-fitting values (X symbols) and, 68\% CIs (lines) of the scaling parameters $\alpha$ and $\beta$ of \autoref{eq:modelab}, for all late-type  galaxies (solid) and their non-AGN subset (dashed). The bias due to the overestimation of SFR in AGN hosts, manifests as an underestimation of the scaling parameter $\alpha$. The value of $\beta$ is highly uncertain to notice any bias due to the inclusion of AGN (lower contribution of old stellar populations in the number of ULXs in LTGs).}
    \label{fig:2dfits}
\end{figure}

As described in \S\ref{txt:observerbias}, the far- and near-infrared emission from AGN may bias the estimates of SFR and \mass{} in our sample. Therefore, when fitting the models in \S\ref{txt:earlylate}, i.e. number of ULXs as a function of SFR and \mass{} in late- and early-type galaxies, we considered only galaxies that were classified as non-AGN in the \hec{}. Here, we perform the same analysis for the complete sample (including the AGN and unclassified galaxies). The results of these fits also enable the direct comparison of this study with previous works where the AGN-hosts were not excluded from their samples.

In the following paragraphs, the complete sample without removing any AGN hosts is referred as \full{} sample. The sample used in \S\ref{txt:results} and \S\ref{txt:discussion}, where galaxies with nuclear activity or lacking classification in the \hec{} were excluded from the fits, is referred as \non{}. The \full{} sample is larger than the \non{} sample by a factor of ${\sim}4$ in elliptical, ${\sim}3$ in lenticular, ${\sim}2$ in spiral and ${\sim1.2}$ in irregular galaxies.

The top panel of \autoref{fig:earlylatefits} shows the scaling of the number of ULXs with the SFR (parameter $a$ in \autoref{eq:modela}) in LTGs (top left) and their morphological sub-classes (top right), for the full (black) and the \non{} (orange) samples. We find $a=0.45^{+0.04}_{-0.03}\aunits{}$ in the case of the \full{} sample, lower than that of the \non{} group, $0.51{\pm}0.06\aunits{}$. However the difference is not significant (${\sim}1\sigma$). The comparison between the \non{} and \full{} sample for the ULX-SFR scaling in different morphological types is not conclusive because of the large uncertainties (see \autoref{fig:earlylatefits}).

The bottom panel of \autoref{fig:earlylatefits} shows the posterior probability distribution of the scaling of the number of ULXs with \mass{} ($b$) for the \full{} (solid line) and \non{} (dashed line) ETGs. We find that the specific ULX frequency is significantly lower when \full{} ETGs are considered ($6.3^{+1.0}_{-0.9}\bunits{}$) than in the case of \non{} ETGs ($15^{+3.9}_{-3.6}\bunits$), with similar results between elliptical and lenticular galaxies.

For the scaling of the number of ULXs with SFR and \mass{} in LTGs, the posterior distribution of the two scaling factors ($\alpha$ and $\beta$) considering the \full{} sample is shown in \autoref{fig:2dfits}. For comparison we also show the \non{} case (dashed lines; presented in \S\ref{txt:earlylate}). We find for the 233 galaxies in the \full{} sample, hosting 328 ULX candidates (48.7 expected f/b sources), $\alpha=0.35^{+0.04}_{-0.05}\sfrunit{}$ and $\beta=6.8^{+2.6}_{-2.2}{\times}10^{-12}\bunits{}$ (marginalised). The difference of the posteriors for the scaling factor $\alpha$ is consistent with the difference seen in the fits with SFR scaling only.

Given their broad-band SED, AGN may have significant contribution in the optical/UV of their host galaxies, and also in their FIR emission in the case of type-2 AGN (cf. \citealt{Risaliti04}). However, their contribution in the near-infrared part of the spectrum is relatively small. Therefore, the difference we find in the specific ULX frequency between the \full{} and \non{} samples of ETGs is unlikely to be due to an overestimation of the $K$-band based \mass{} estimates in possible AGN in the full sample.

This is supported by the X-ray luminosities of the nuclear sources in the full ETG sample. These are very low (only one exceeds $10^{41}\ergs{}$), indicating that if there is an AGN its contribution to the $K$-band luminosity may not be significant. Additional support comes from the ratio of the nuclear and the total $K$-band luminosity in a representative sample of star-forming galaxies of Bonfini et al. (2020; submitted). This study (based on morphological decomposition of $K$-band images of the Star-Formation Reference Survey; \citealt{Ashby11}) shows that the typical AGN contribution to the galactic $K$-band luminosity is ${<}10\%$. We conclude that the \mass{} estimates in AGN-hosts are not expected to be significantly biased upwards by the potential presence of an AGN.

Therefore, as discussed in \S\ref{txt:disearly}, the difference in the specific ULX frequency in the non-AGN and the \lq{}full\rq{} ETG samples is the result of the SFH differences in ETGs of different stellar masses.

\section{Detailed comparisons with previous ULX surveys}
\label{app:comparisons}

In the following subsections we compare our results with two major surveys of ULXs in the local Universe: \citet{Swartz11} (hereafter \citetalias{Swartz11}) and \citet{Wang16} (hereafter \citetalias{Wang16}).

\subsection{Comparison with \textit{Swartz et al. (2011)}}
\label{app:swartz}

Based on the total LTG sample in our study, we find a SFR scaling factor of ${\sim}0.51{\pm}0.06\aunits{}$ (see \autoref{tbl:earlylate}) which is ${\sim}4$ times lower than that found by \citetalias{Swartz11} (${\sim}2\aunits{}$).

The main reason behind this discrepancy is the different samples of host galaxies in terms of morphology: \citetalias{Swartz11} consider a large population of late-type and irregular galaxies, as shown in their fig.~1). Using the reported fractions of morphological types in their sample, and our fitting results for $a$ in different sub-classes of LTGs (see \autoref{tbl:earlylate}), we compute the expected number of ULXs in a sample with the same distribution of morphological types as in \citetalias{Swartz11}, based on our findings. We find $1.2\aunits{}$, a factor of 2.4 higher than in our sample. Another reason for this discrepancy could be differences in the X-ray photometry of the sources\footnote{\citetalias{Swartz11} used a count-rate to flux conversion factor assuming an absorbed power-law model with $\Gamma=1.8$ and, for sources with $>130$ counts, performed spectral fits or adopted published results. Instead, we adopt the aperture-corrected net energy flux from the \CSC{}.}. Indeed, for the 99 common X-ray sources (cross-match radius of 3\arcsec{}), only 73 of them are characterised as ULX candidates by us, while we find that the luminosities we report are smaller by ${\sim}-0.4\rm\,dex$ (a factor of ${\sim}0.4$) compared to those computed by \citetalias{Swartz11}. Assuming a cumulative slope of $0.6$ (appropriate for the HMXB-LF; \citealt{Grimm03,Mineo12}), this corresponds to a lower number of ULXs by a factor of ${\sim}1.7$ compared to \citetalias{Swartz11}. In total, the combination of the two factors give a factor of $4.2$ lower estimate of the SFR scaling factor in our sample, explaining the difference we find.

Finally, in our sample, we find ${\sim}0.68\pm0.10$ ULXs per elliptical galaxy (see \autoref{tbl:ulxrates}) which is significantly higher than the rate (0.23) reported in \citetalias{Swartz11}. We attribute this discrepancy to the small number statistics, and the under-representation of elliptical galaxies in \citetalias{Swartz11}. On the other hand, the fraction of elliptical galaxies in our sample is similar to that of \citetalias{Wang16}, which presents comparable number of ULXs per elliptical galaxy ($0.43{\pm}0.11$).

\subsection{Comparison with \textit{Wang et al. (2016)}}
\label{app:wang}

We cross-check the number of ULXs in our sample against the results of \citetalias{Wang16}, the largest and most recent study of ULX demographics with \chandra{} observations. To do so, we compare the total number of ULXs in the common galaxies in our sample and in that of \citetalias{Wang16}.

As a first step, we cross-match the two galaxy samples. Out of the 343 galaxies in the sample of \citetalias{Wang16}, 315 are associated with our host galaxy sample. The remaining 28 galaxies are not included in our sample for various reasons. In 22 cases, the targets were observed with shallow observations (exposure times ${\lesssim{}}5\,\rm ks$) and \chandra{} did not detect any source. In the remaining six cases, the sources in \citetalias{Wang16} do not lie in the $D_{25}$ regions of the galaxies, our criterion for associating sources to host galaxies (NGC3066, NGC1507), or the observations were not included in the \CSC{} (NGC3489, NGC3489, PGC48179, PGC35286).

However, there are three additional important differences between this study and that of \citetalias{Wang16} that need to be considered in the comparison:
\begin{enumerate}
    \item In \citetalias{Wang16}, X-ray sources must exceed $2{\times}10^{39}\ergs{}$ in X-ray luminosity to be considered as ULX candidates, instead of our criterion of $L_{\rm X}{>}10^{39}\ergs{}$.
    \item The X-ray source sample of \citetalias{Wang16} was taken from \citet{Liu11} who computed the fluxes of the sources using a count-to-flux conversion assuming a power-law spectrum with $\Gamma{=}1.7$ and Galactic line-of-sight absorption. Instead, we use the net energy of the photons as reported in the \CSC{} (see \S\ref{txt:luminosities}).
    \item In this study, we associate sources to a host galaxy if they lie in its $D_{25}$ region. In \citetalias{Wang16} the $2{\times}D_{25}$ regions are used, namely the sky ellipses with twice the major and minor axes.
\end{enumerate}

In order to account for the luminosity difference, we cross-match our X-ray sample with the one used by \citetalias{Wang16}, and we find that our luminosities are $6\%$ smaller (median ratio; scatter of $0.5\rm\,dex$). Therefore, for this comparison only, we will consider as ULXs in our sample, the \good{} X-ray sources with $L_{\rm X} >1.89\times10^{39}\ergs{}$ to account for the above mentioned luminosity offset. We find a total of 186 ULX candidates in our sample. The same luminosity limit is used to calculate the expected foreground/background contamination (see \S\ref{txt:interlopers}), for which we find 37.9 sources, leading to a final estimate of 148.1 ULXs in the 315 galaxies, based on our emulation of the \citetalias{Wang16} sample.

To correct for the difference between the number of sources in the $D_{25}$ and $2{\times}D_{25}$ regions, we cross-match the X-ray sample with the \hec{} and find that the number of X-ray sources that lie in the $2{\times}D_{25}$ regions is $35\%$ larger than the number of sources in the $D_{25}$ regions. Therefore, the number of ULX candidates in the $2{\times}D_{25}$ regions, 215, as reported in \citetalias{Wang16}, corresponds to about 159.3 candidates in the $D_{25}$ regions. By subtracting one forth of the, reported by \citetalias{Wang16}, number of interlopers (since the area of the $D_{25}$ regions is four times than the area of the $2{\times}D_{25}$ regions), we find that the number of ULXs in the $D_{25}$ regions is 151.1, close to the value we find in our sample (148.1).

Lastly, the ULX rates above 2, 3, 5 and 10$\times10^{39}\ergs{}$ in the different morphological types in our sample (see \autoref{tbl:ulxrates}) are consistent at the $2\sigma$-level with the rates found in the table~2 in \citetalias{Wang16}.


\bsp	
\label{lastpage}
\end{document}